# In-vivo real-time $^{13}$C-MRSI without polarizer on site: across cities transportable hyperpolarization using UV-induced labile radicals

Andrea Capozzi [1,2 *], Magnus Karlsson [2], Yupeng Zhao [2], Jan Kilund [2], Esben Søvsø Szocska Hansen [3], Lotte Bonde Bertelsen [3], Christoffer Laustsen [3], Jan Henrik Ardenkjær-Larsen [2], and Mathilde H. Lerche [2].

[1] *LIFMET, Department of Physics, EPFL, Station 6 (Bâtiment CH), 1015 Lausanne (Switzerland).*

[2] *HYPERMAG, Department of Health Technology, Technical University of Denmark, Building 349, 2800 Kgs Lyngby (Denmark).*

[3] *The MR Center, Department of Clinical Medicine, Aarhus University, Palle Juul-Jensens Boulevard 99, 8200 Aarhus N (Denmark)*

**Corresponding author**

[*] Dr. Andrea Capozzi

ORCID: 0000-0002-2306-9049

EPFL SB IPHYS LIFMET

CH F0 632 (Bâtiment CH), Station 6

CH-1015 Lausanne

Email: andrea.capozzi@epfl.ch

## Abstract (150 words)

Hyperpolarized $^{13}C$ Magnetic Resonance Spectroscopic Imaging (HP $^{13}C$-MRSI) has the potential to greatly improve diagnostic radiology thanks to its unique capability to detect, noninvasively, a wide range of diseases entailing aberrant metabolism. Nevertheless, it struggles to enter everyday clinical practice as an alternative and/or complement to Positron Emission Tomography (PET). Because of the minute-long hyperpolarization lifetime of the MR HP contrast agents, one of the reasons is, differently from PET, the need and financial burden to have the "polarizing machine" on site and, as close as possible to the MR scanner.

In this work, we show that dDNP samples prepared with UV-induced labile radicals can bridge the technical gap with PET and provide MRI facilities with hours-long relaxing HP contrast agents. As a demonstration, we show the first "across cities" HP $^{13}C$-MRSI experiment *in vivo* and on a clinical scanner for a perfusion/angiography ([1-$^{13}C$]HP001) and a metabolic ([U-$^{13}C$, $d_7$]glucose) contrast agent.

## Manuscript (5000 words)

## Introduction

According to the World Health Organization, diagnostics are the most important tools that empower the health workforce in the identification of diseases. They allow the initiation of treatments for patients in order to avoid further complications, improve prognosis, and reduce side effects and expenses related to more invasive protocols, later on.

Some severe conditions like cancer, diabetes, neurodegenerative diseases and cardiovascular diseases are initiated by metabolic changes[1]. Therefore, imaging methods able to investigate metabolic processes *in vivo* can represent a wealth of information concerning the origin, kind and severity of the metabolic aberration[2–5]. Currently, the clinical benchmark to investigate hypo- and hypermetabolism is $^{18}$F-Fluoro-Deoxy-Glucose positron Emission Tomography ($^{18}$F-FDG-PET). This technique, although routinely used to asses cancer diagnosis, staging and treatment monitoring[6], it has its downsides. Besides poor specificity in organs characterized by normal high glucose uptake like the brain, and the employment of ionizing radiations that limit its use for repeated monitoring and in certain patients groups[7], most importantly, $^{18}$F-FDG-PET falls short to discern a metabolic product from a substrate and access the rich information behind metabolic fluxes[8]. Indeed, FDG is, like glucose, taken up by living cells through glucose transporter membrane proteins and subsequently phosphorylated by the enzyme hexokinase. But, contrary to phosphorylated glucose, phosphorylated FDG does not metabolize further, which makes it an effective biomarker for glucose uptake, but prevents it from reporting directly on glycolytic activity (downstream glucose metabolism).

Magnetic Resonance Spectroscopic Imaging (MRSI) is an ionizing radiation free technique with the great advantage of chemical shift: substrate and metabolites resonate all at slightly different frequencies, making the spectrum of each molecule its peculiar and, especially for $^{13}C$, clear fingerprint. Nevertheless, compared to PET, it pays the heavy toll of low sensitivity[9].

Hyperpolarization methods[10,11], by generating *ex-situ* isotopically labeled and injectable endogenous compounds with a MR signal enhancement of $10^4 – 10^5$ orders of magnitude (i.e. HP MR contrast agents), gifted $^{13}C$ MRSI with unprecedent time and spatial resolution[12,13], giving access to metabolic fluxes in real-time and *in-vivo*.

As personalized oncology becomes a reality, there is a requirement to diagnose tumors and detect their response to treatment earlier, better determine their aggressiveness and prognosis, and identify novel treatment's targets. $^{13}C$ HP MRSI has the potential to do so in a non-invasive manner[14], and over the past two decades has gained great significance in preclinical studies thanks to three hyperpolarization techniques: PHIP-SAH (ParaHydrogen-Induced Polarization-Side Arm Hydrogenation)[15,16], SABRE (Signal Amplification by Reversible Exchange)[17], and dDNP (dissolution Dynamic Nuclear Polarization)[18]. So far, thanks to its versatility (no requirement for substrate molecule specific optimization)[19], straightforward biocompatibility (no employment of organic solvents and metal complex catalysts), and superior polarization level at injection[20–22], only dDNP made it into humans with the molecule [1-$^{13}C$]pyruvate to diagnose cancer and detect treatment response[23–25], and [$^{13}C$, $^{15}N$]urea soon to come for tumor perfusion purposes.[26]

Nevertheless, this methodology struggles to enter everyday clinical practice as complement and/or alternative to PET. One of the reasons why broad consensus among clinicians is still missing lies in the technical complexity that characterizes hyperpolarization via dDNP[27]. A main drawback is that, differently from PET, dDNP HP compounds prepared employing traditional Electron Polarizing Agents (EPA), i.e. stable organic radicals, have to be produced and turned into injectable solutions on site because of their minute-long half-life in the liquid state[28]. Currently, should you be willing to equip an MRI facility with hyperpolarization, the only way is to place, as close as possible to the scanner, costly and technically demanding hardware (i.e. the dDNP polarizer). Lifting this restriction and making hyperpolarization transportable would allow centralized substrates hyperpolarization followed by transport to multiple locations for clinical use, in a similar fashion to $^{18}$F-FDG PET tracers, dispensing with the requirement to have a dDNP polarizer in close proximity[14].

The half-life of $^{13}$C labeled MR contrast agents is orders of magnitude longer when kept at cryogenics temperatures[29,30]. Nonetheless, making hyperpolarization transportable has one caveat: to limit or eliminate paramagnetic relaxation of the nuclei of interest at the moment of extraction of the HP substrate as a frozen solid from the dDNP polarizer. Indeed, the same EPA used, upon microwave irradiation, to create the HP state in the first place, would relax the high spin order within a few milliseconds outside the high magnetic field of the dDNP polarizer[31]. So far, to circumvent this shortcoming, researchers have pursued two different approaches:

- To use UV-induced labile radicals that can be quenched, after the DNP process has happened in the solid state inside the polarizer, by heating up the sample above 190–200 K by means of a thermalization process[32–35];
- To physically separate carbon nuclei from the polarizing agents by playing with radicals and $^{13}$C-labelled molecules with different solubility, or by grafting the radicals inside purpose synthesized porous polymers that can absorb a solution containing the substrate of interest[36,37].

HP substrates polarized with UV-induced labile radicals have the advantage of being readily biocompatible, because the radicals' precursors belong to the alpha-keto acids family, most of which are endogenous products. Moreover, the natural radical scavenging upon temperature increase of the sample removes the need for filtration and associated quality control required for clinical applications when synthetic radicals are used[12].

As of today, all attempts of transport have been limited to a very small scale and to the level of proof-of-principle *in vitro* experiments. In this paper, combining UV-induced labile radicals, a robust strategy for sample polarization, thermalization, extraction and storage[38], we demonstrate the first "across cities" HP $^{13}$C-MRI experiment *in vivo* and on a clinical scanner for a perfusion/angiographic ([1-$^{13}$C]HP001) and a metabolic ([U-$^{13}$C, $d_7$]glucose) HP MR contrast agent.

## Results

In Figure 1, we summarize the Electron Spin Resonance (ESR) and solid-state DNP properties of the reference sample of this study: a solution of 2 M [U-$^{13}$C, $d_7$]glucose (from now on referred as glucose) and 0.7 M alpha-ketoglutaric acid (from now on referred as AKG) dissolved in $D_2O$:$d_8$-glycerol 1:1 (v/v) (from now on referred as deuterated solvent). In Figure 1A, we used X-band ESR spectroscopy to quantify the radical generation upon UV-light irradiation in liquid nitrogen[39] of 15 x 10 μL frozen beads of sample's solution (see Material and Methods for details). A UV-irradiation at 35 W/cm$^2$ was sufficient to saturate the radical generation around 20 mM after 300 s, with a quantum yield of almost 3%. The inset shows the derivative of the ESR spectrum at the end of the irradiation. The latter was mainly characterized by the dipolar coupling of the unpaired electron localized on the secondary carbon atom with the adjacent 2H group[32]. As earlier demonstrated, radical generation followed a mono-exponential growth with a time constant of 29±1 s ($R^2$ = 0.95)[39]. In Figure 1B, the radical quench temperature was measured using the X-band ESR spectrometer equipped with a Variable Temperature Insert (VTI). The result was a steep drop in the radical concentration between 175 K and 200 K. No ESR signal could be detected above 225 K (see Figure S1 in Supporting Information). Figure 1C shows the ESR and DNP properties of the sample at 1.2 K and 6.7 T. The high field ESR spectrum, measured by means of Longitudinal Detection technique [39,40], was overlayed to the $^{13}C$ NMR signal as a function of the microwave irradiation frequency, i.e. the DNP spectrum. The latter was measured twice by keeping all experimental parameters unchanged except for the microwave frequency output, kept constant in the first case (mono-chromatic irradiation), and swept at a rate of 1 kHz around the central frequency by ±25 MHz (50 MHz of microwave frequency modulation) in the second. Both DNP spectra nicely overlapped with the ESR one, showing that the enhancement mechanism happened only for microwave frequencies where electron spins were also resonating, a clear sign of DNP by Thermal Mixing or Cross-Effect[41,42]. More interestingly from a practical standpoint, the DNP signal increased by 4-fold when microwave frequency modulation (FM) was used. To further investigate this observation, we measured, by means of LOD-ESR the electron spin dynamic of the EPA in correspondence of the non-modulated DNP positive maximum (i.e. 188.02 GHz) for increasing FM values (see Figure 1D).

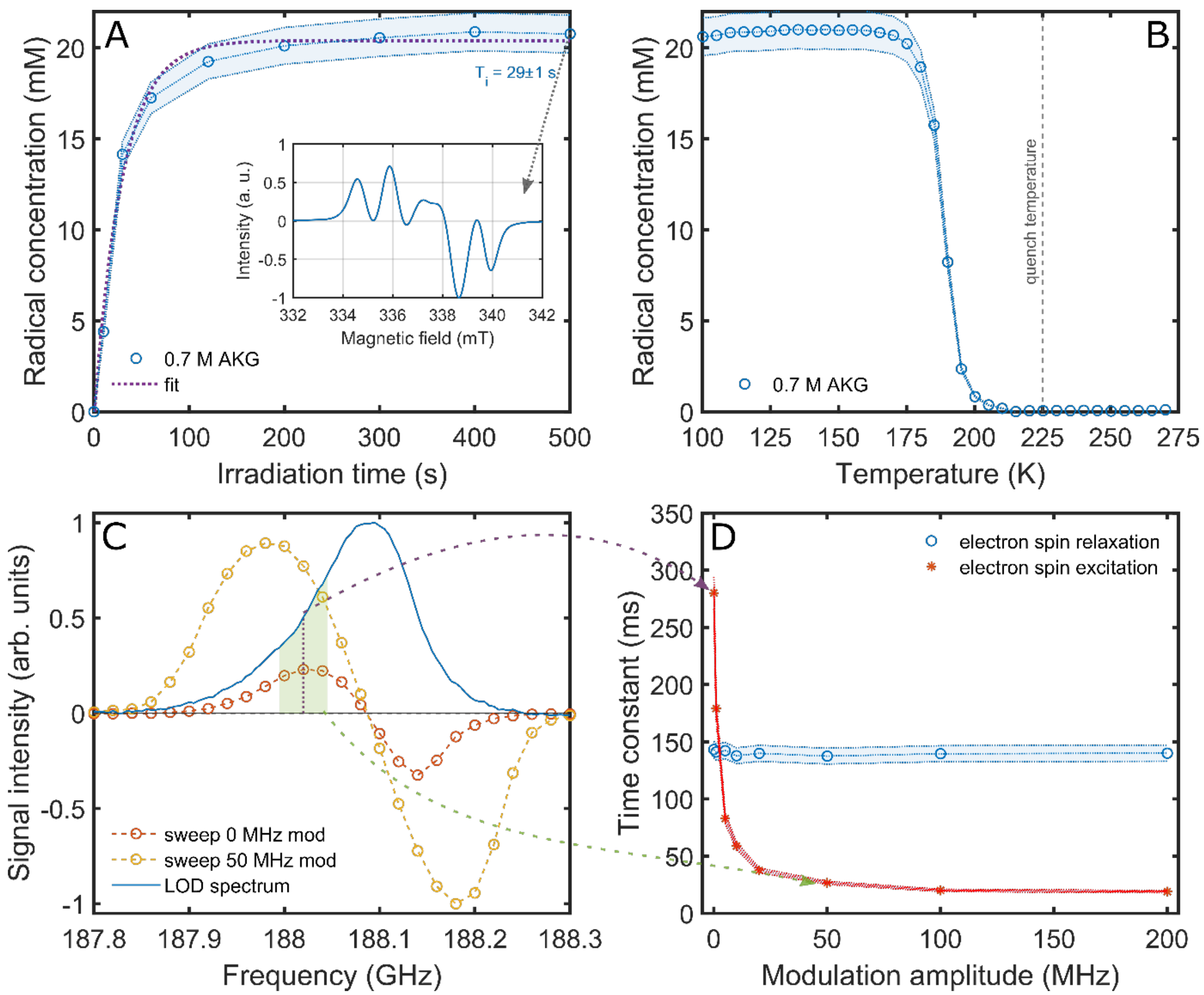

**Figure 1: ESR and DNP properties of the sample.** Radical generation measured by means of X-band ESR spectroscopy at 77 K in a small quartz bath cryostat; 15 x 10.0±0.5 μL frozen beads of sample's solution were irradiated in liquid nitrogen with high power (35 W/cm$^2$) UV-light for increasing time intervals, and measured via ESR at the end of each interval; a mono-exponential curve (violet dotted line) was fit to the data; the inset shows the 1$^{st}$ derivative of the ESR spectrum corresponding to the last data point of the time-course (**A**). Radical's quench temperature measured by means of X-band ESR spectroscopy; the ESR spectrometer was equipped with a Variable Temperature Insert (VTI) and one bead was transferred inside the sensitive area of the VTI after complete UV-irradiation; the temperature was increased in steps of 5 K and the ESR signal measured after thermal stabilization of the system; the grey dashed line indicates when no more signal could be measured from the sample (**B**). Sample's DNP and LOD-ESR spectra at 6.7 T and 1.2 K; after loading the sample inside the dDNP polarizer, its ESR spectrum at high magnetic field was measured by means of the LOD technique (blue line); the $^{13}$C signal enhancement as a function of the microwave irradiation frequency (i.e. the DNP spectrum) was also measured with (yellow circles) and without (orange circles) microwave frequency modulation; the positive DNP maximum (violet dotted line) and the portion of irradiated ESR spectrum corresponding to 50 MHz of frequency amplitude modulation (green rea) are also indicated (**C**). To investigate the effect of microwave frequency modulation on the radicals' electron spins, their dynamic upon excitation (microwaves go ON, red circles) and relaxation (microwaves go OFF, blue circles) was also measured for increasing values of the microwaves modulation amplitude (**D**). In all panels, errors on the measurements are reported as shaded areas around the data points.

While the characteristic time constant of the electron spins system upon turning OFF the microwaves (i.e. electron spins relaxation, $T_{eR}$) stayed constant around 150 ms, the dynamic upon turning ON the microwaves (i.e. electron spins excitation $T_{eE}$) decreased from 280±15 ms, for monochromatic irradiation, to 25±2 ms, for 50 MHz of FM. No further decrease was observed for higher values of FM. It is worth noting that we prefer to talk about relaxation and excitation instead of the commonly used "electron $T_1$" and saturation constant because the dynamic we observed is mixed with spectral diffusion and we actually don't know whether or not we reach full saturation of the ESR line.

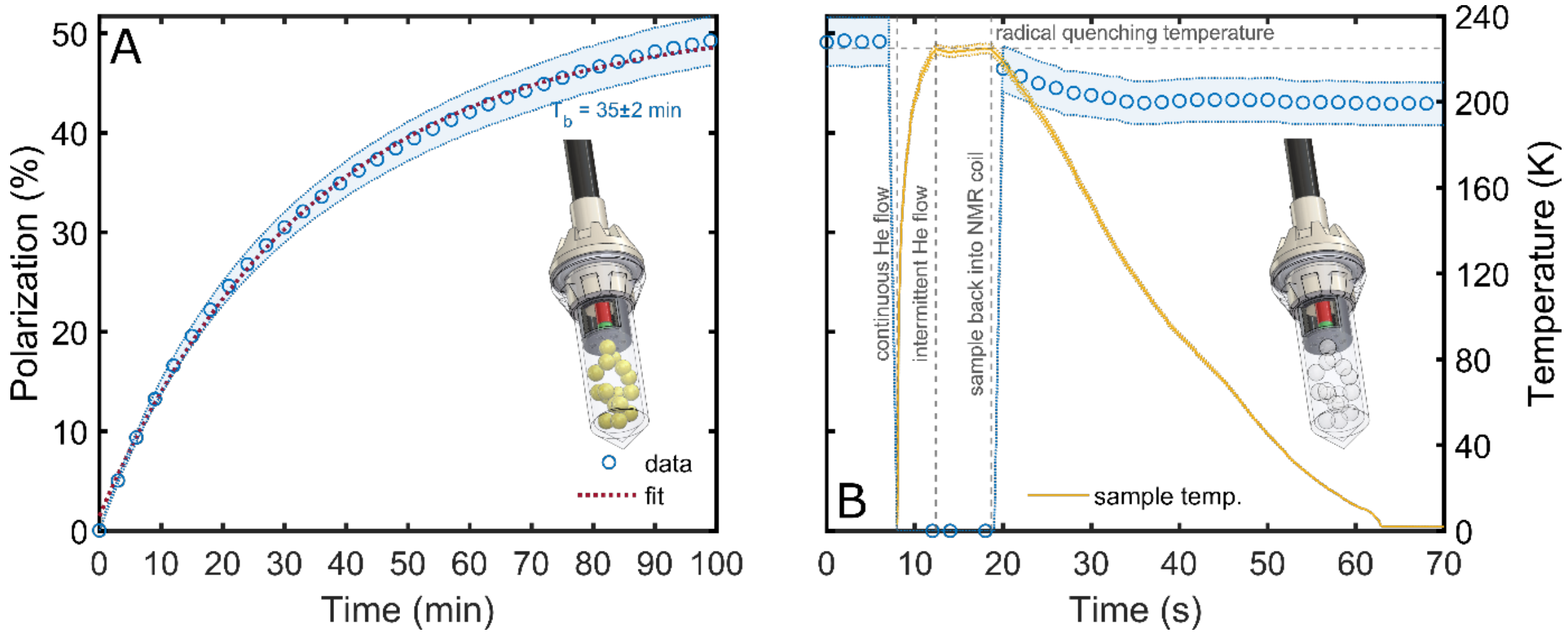

**Figure 2. Sample's polarization and thermalization.** $^{13}C$ signal time-course (blue circles) at best microwave irradiation conditions (frequency = 188.18 GHz, frequency amplitude modulation = 50 MHz); a mono-exponential

curve (violet dotted line) was fit to the data points (**A**). $^{13}C$ signal time-course (blue circles) during sample thermalization with overlayed temperature profile (yellow line); the vertical dashed grey line indicate the different steps of the thermalization process; the horizontal dashed grey line indicates the set temperature threshold during heating by He gas blowing (**B**). The two insets show the sample appearance at the bottom of the CFP (see text for details) before (colored frozen beads) and after (transparent frozen beads) the thermalization process; the color loss was due to the recombination of radicals into diamagnetic species. In both panels, errors on the measurements are reported as shaded areas around the data points.

Figure 2 shows the sample's polarization buildup and thermalization above the radical quenching temperature. A Custom Fluid Path (CFP)[34,40], equipped with a temperature sensor, was used to seal and load the sample inside the dDNP polarizer, and monitor its heating profile during thermalization. The sample was first polarized at best microwave conditions (carrier frequency 188.05 GHz , output power 50 mW, FM 50 MHz) obtaining a solid state $^{13}C$ polarization of 48±2 % after 90 min (panel A). The left inset displays a sketch of the bottom part of the CFP (i.e. what is immersed in liquid He and sits inside the NMR coil) during polarization. The sample's beads are colored to indicate the presence of radical.

After polarization, the sample was thermalized using room temperature He gas (panel B). This procedure allowed us to get rid of more than 99 % of the initial radical's concentration (see Figure S2 in Supporting Information) in a reproducible and robust manner, by limiting polarization losses to no more than 20 % of the maximum value achieved during DNP. The right inset displays the sample as transparent beads because the thermalization procedure and consequent radical annihilation made them lose their characteristic color.

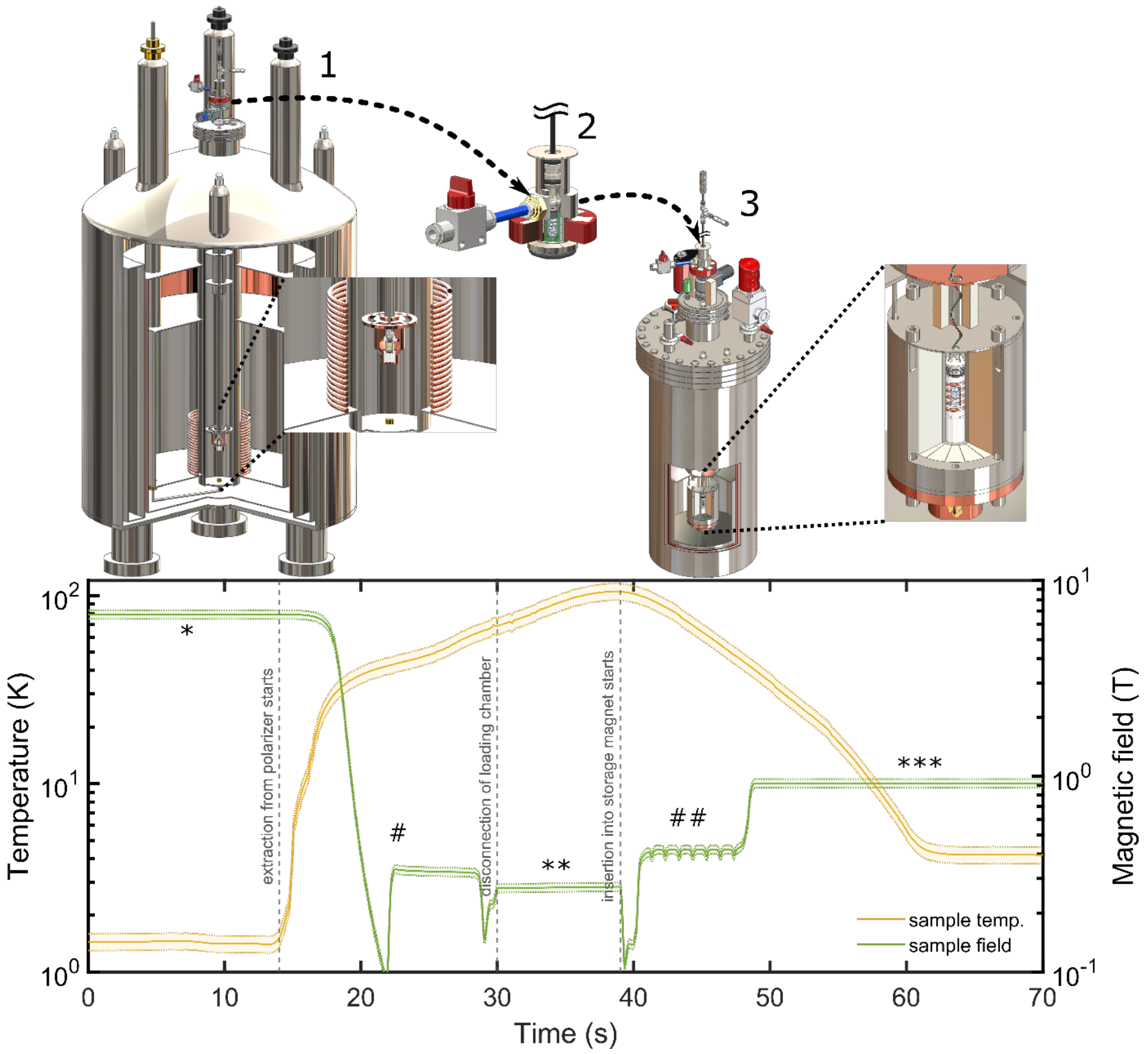

**Figure 3. HP sample's extraction and storage procedure.** Sketch of sample's extraction and storage step-by-step procedure: (1) after thermalization, the FP is manually pulled through the dynamic seal and the vial containing sample lifted, along the magnetically enforced probe of the polarizer, up to the loading chamber; (2) the gate valve is closed and the loading chamber disconnected from the polarizer and docked to the portable cryostat; (3) the gate valve of the transportation device is opened and the vial containing the sample pushed, along the magnetically enforced probe of the portable cryostat, down to the liquid He cooled permanent magnet (**top panel**). Magnetic field (green line) and temperature (yellow line) experienced by the sample during these operations; errors on the measurements are reported as shaded areas around the data points; the vertical dashed grey lines indicate the onset of each of the three main steps; special characters indicate regions of "constant" magnetic field: (*) polarizer superconductive coil; (#) Halbach array

along the polarizer probe; (**) Halbach array around loading chamber; (##) Halbach array along the portable cryostat's probe; (***) 1 T NMR quality permanent magnet (**bottom panel**).

The extraction from the polarizer (Figure 3, top panel) into a purpose built transportable liquid He cryostat with NMR capability [38] took approximately 50 s. The temperature and magnetic field experienced by the sample during this procedure are reported in Figure 3 (bottom panel). It is important to notice that different arrangements of permanent magnets inside the polarizer and the portable cryostat allowed us to always keep the sample in a magnetic field of at least 100 mT. Details concerning the design of the magnetically enforced probe of the dDNP polarizer and portable cryostat were described earlier [34,38].

During this process, the sample's temperature quickly increased from 1.2 K, at the beginning of the extraction to approximately 30 K when the vial was raised to the loading chamber; displacing the loading chamber from the polarizer and docking it at the top the portable cryostat took approximately 20 s and generated a further temperature increase to 100 K, but at a lower rate; once inside the portable cryostat, it took further 20 s to cool the sample down to 4.2 K.

It is worth noting that transferring the HP sample from the transportable cryostat back inside the DNP polarizer showed that, withing 5% error, the signal intensity was equal to the one from the last acquisition before extraction, proving the procedure to be basically loss free.

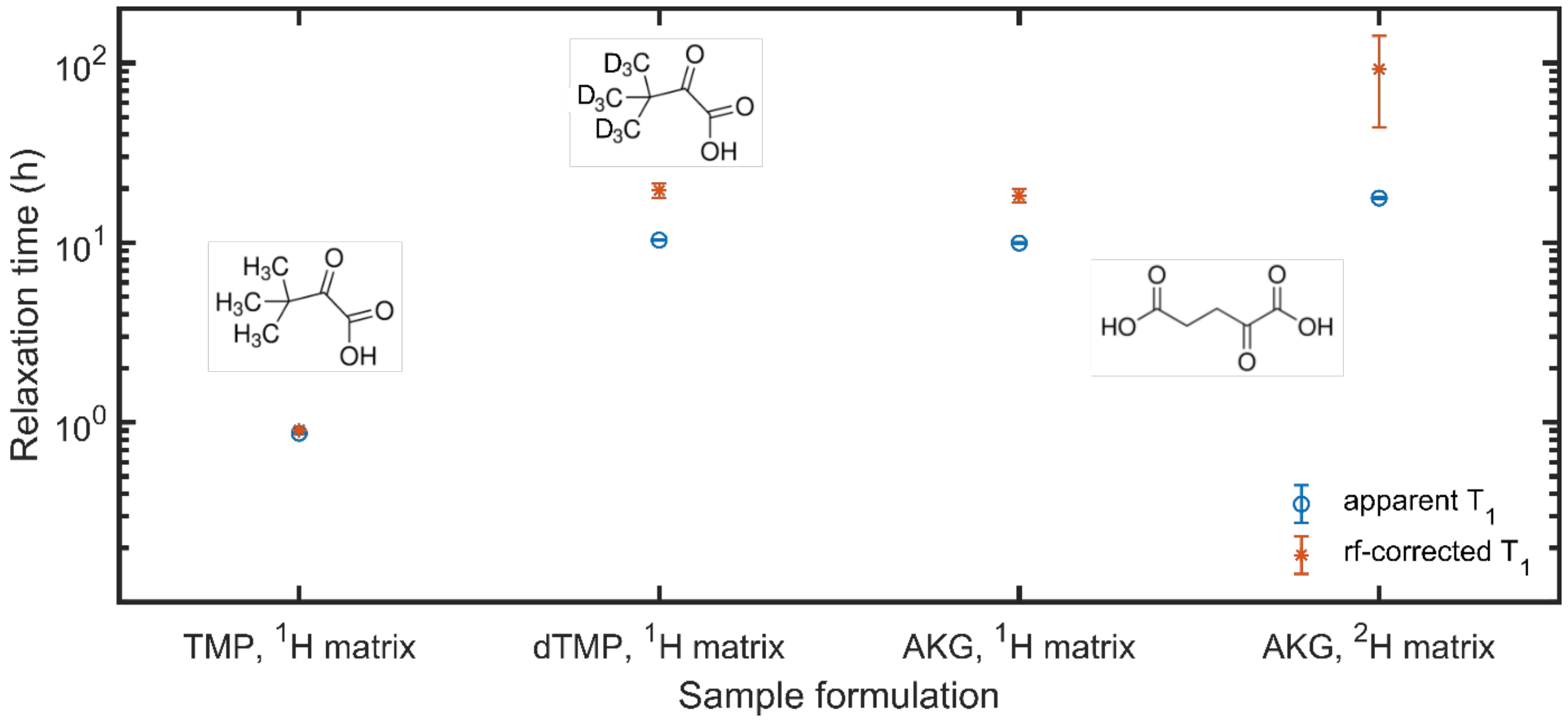

**Figure 4. Relaxation at storage conditions.** $^{13}C$ nuclei solid-state relaxation time constant at 1 T and 4.2 K for samples containing 2M [U-$^{13}C$, $d_7$]glucose that have been polarized, thermalized, extracted and stored inside the portable cryostat as a function of the molecular structure of the radical precursor and the deuteration of the glassy matrix of the solvent. Four formulation are reported: 0.7 M of trimethylpyruvic acid dissolved in $H_2O$:glycerol 1:1 (v/v) (TMP, $^1H$ matrix); 0.7 M of $d_9$-trimethylpyruvic acid dissolved in water:glycerol 1:1 (v/v) (dTMP, $^1H$ matrix); 0.7 M of alpha-ketoglutaric acid dissolved in $H_2O$:glycerol 1:1 (v/v) (AKG, $^1H$ matrix); 0.7 M of alpha-ketoglutaric acid dissolved in $D_2O$:$d_8$-glycerol 1:1 (v/v) (AKG, $^2H$ matrix).

Polarization, extraction and storage were repeated on different glucose sample preparations in terms of radical precursor's molecule and level of deuteration of the solvent (i.e. glycerol:water 1:1 (v/v)) (see Figure 4). While minimal differences were measured regarding radical yield, quench temperature, and DNP performance (see Capozzi et al[32], and Figure S3 and Figure S4 in Supporting Information), the impact of these 2 parameters had a huge impact on the relaxation time value at storage conditions. For the same radical concentration's residual after thermalization (approximately 0.5% of the initial value, within experimental errors), the glucose $^{13}C$ $T_1$ of the sample at 4.2 K and 1 T measured 0.87±0.01 h for trimethylpyruvic acid (TMP) as radical precursor and in a protonated solvent, 10.3±0.1 h for d4-trimethylpyruvic acid (dTMP) as radical precursor and in a protonated solvent, 9.9±0.1 h for AKG as radical precursor and in a protonated solvent, and 17.7±0.2 h for AKG as radical precursor and in a deuterated solvent. These values increased to 0.90±0.01 h, 19.5±1.8 h, 18.2±1.5 h, and 92.4±48.8 h correcting for the effect of the NMR measuring pulse, respectively (see Figure S5 and Figure S6 in Supporting Information).

The sample formulation providing the longest relaxation time (2 M glucose and 0.7 M AKG dissolved in $d_8$-glycerol:$D_2O$) was used to perform HP MRSI *in-vivo*. Thanks to the transportable cryostat's holding time of up to 8 h[38], we were able to hyperpolarize, thermalize, extract and store the sample at HYPERMAG (Technical University of Denmark, Kgs Lyngby, Denmark) and transport it by car to the MR Center at Aarhus University Hospital (Aarhus, Denmark), where a healthy rat had been prepared in a 3T GE clinical scanner, and anatomical images already acquired. The trip took 5 h and we covered 320 km. Once we arrived on site, all we had to do was to place the transportable cryostat in front of the entrance of the scanner room and power up the dissolution unit to quickly melt the HP solid sample.

After dissolution, 1 mL of a 60 mM HP glucose solution was manually transferred into the scanner room, injected through a tail vein catheter, and the Chemical Shift Imaging (CSI) acquisition sequence started with the observation frequency centered on the c2-c5 glucose multiplet. Considering losses during transport, an elapsed time of 30 s between dissolution and beginning of the MR acquisition, and a glucose liquid-state $T_1$ of 15 s[34], we estimated a $^{13}C$ polarization between 5 and 7 % at injection. As a first attempt and because of the broad chemical shift of glucose we used all the HP signal to acquire one single image (see Figure S7 in Supporting Information)

The experiment was repeated hyperpolarizing a compound with better properties for HP imaging: HP001(bis-1,1-(hydroxymethyl)-[1-$^{13}C$]cyclopropane-$d_8$). The DNP sample was prepared in the same way, but the 2M glucose was replaced by 5 M of HP001. After polarization, thermalization and extraction the $^{13}C$ HP signal from the transport cryostat was observed for 45 min prior to acknowledging negligible losses at storage conditions (see Figure S8 in Supporting Information). Before our arrival at Aarhus University Hospital, a healthy rat had been imaged already hyperpolarizing HP001 on site and using trityl radicals and a SpinAligner polarizer [43,44]. The liquid-state signal of a 150 mM single peak $^{13}C$ solution was used to acquire a complete angiogram by means of a gradient echo sequence with frames every 0.5 s. Figure 5A clearly shows the rat cardiovascular system's highlight starting from the tail vein injection. Complete coverage of heart and kidneys was achieved at frame #9 (Figure 5B), after 4 s from injection (the first frame had no signal because the acquisition was started few instants before injection). Afterwards, the signal slowly started to fade out.

The hyperpolarized angiogram was repeated on a second rat using the transported hyperpolarized signal (Figure 5C). This time, complete coverage of heart and kidneys was achieved at frame #8 (Figure 5D). For both experiments, 1 mL was used for the tail vein injection. A second syringe was brought to a benchtop NMR spectrometer to estimate polarization at injection. With a $T_1$ of 82±2 s (measured at 1 T on the benchtop spectrometer) and a transfer time of 30 s the $^{13}C$ polarization at injection was estimated to be 52% and 24% for the polarized-on-site and transported sample, respectively.

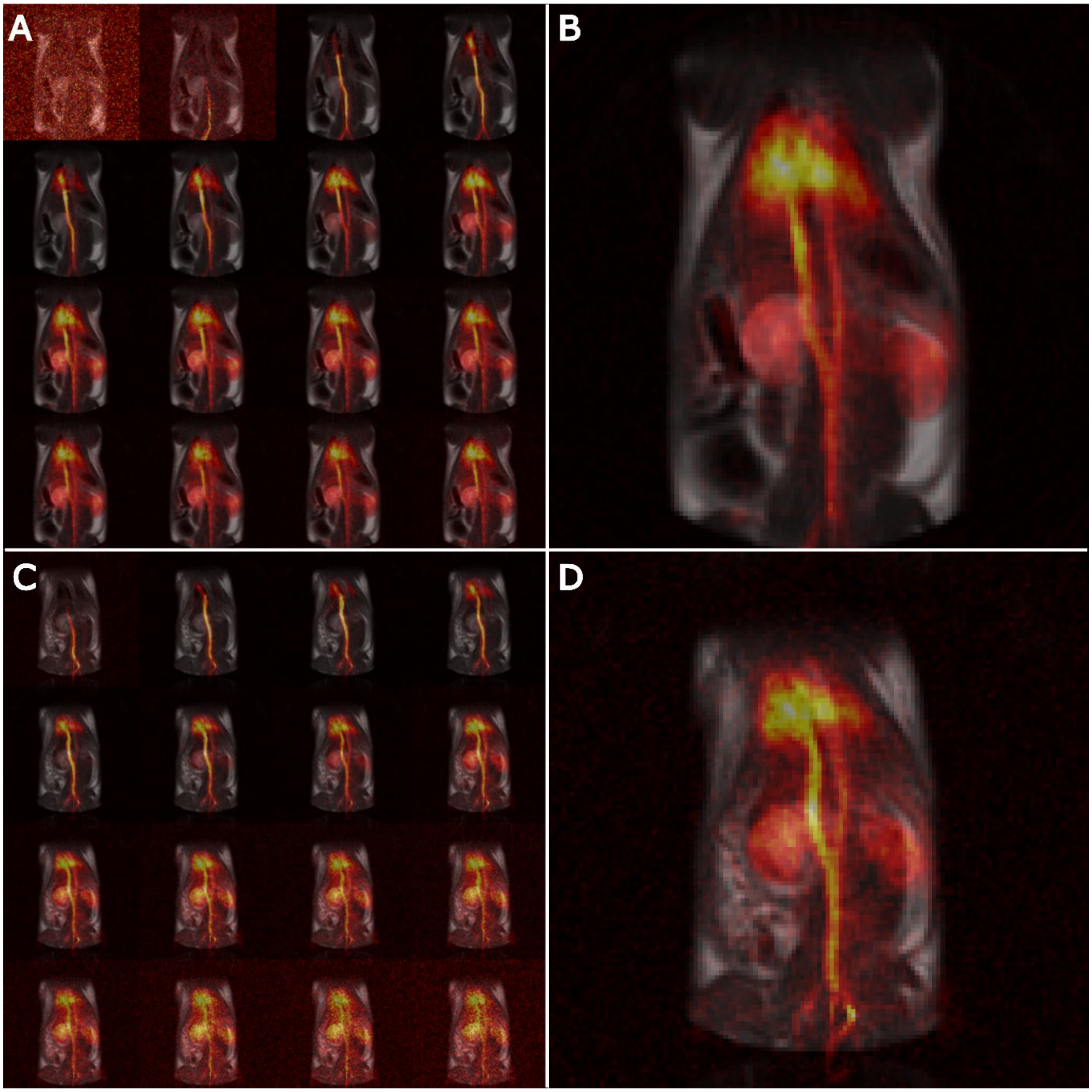

**Figure 5. Rat angiography using hyperpolarized HP001(bis-1,1-(hydroxymethyl)-[1-13C]cyclopropane-d8)**. The first rat was injected with the substrate polarized on-site, at Aarhus University Hospital, and using trityl radicals; polarization at injection was approx. 54%. In the time series one image was acquired every 0.5 s (**A**). A zoom of frame #9, corresponding to the max coverage of the cardiovascular system before the signal starts fading out, is also reported (**B**). The second rat was injected with the substrate polarized at DTU, thermalized to make it "radical free" and long relaxing, and transported over 320 km for 5 h to Aarhus University Hospital; polarization at injection was approx. 24%. In the time series one image was acquired every 0.5 s (**C**). A zoom of frame #8, corresponding to the max coverage of the cardiovascular system before the signal starts fading out, is also reported (**D**).

## Discussion

Our results clearly show that we pushed transportable hyperpolarization well beyond proof-of-concept. A thorough handling of all key parameters (i.e. high enough radical concentration, efficient $^{13}C$ DNP above 50% polarization, minimal losses during thermalization, extraction and storage) allowed us to perform HP MRSI *in-vivo* and on clinical scanner with a contrast agent polarized off-site. The achieved level of contrast in the images was not far from what was obtain from a standard trityl sample polarized on-site. Half of the polarization was lost during the full experimental procedure. Yet, > 20% carbon polarization represents the clinical threshold for a HP $^{13}C$ MRSI to be meaningful (Spinlab ref???).

If HP transportation aims at covering a shorter distance, liquid nitrogen can be used. At 77 K and 1 T, a carbon $T_1$ of more than 30 min was measured on the reference sample (see Figure S11), with the advantage that, at these experimental conditions, the transportable cryostat can be half the size, the weight and cost.

A very important outcome of this study is shown in Figure 4. Here, differently from what was explained in the introduction, we clearly show that to make the high spin order of a sample polarized via DNP lasting for hours, moderate magnetic field, low temperature and radicals' absence/negligible residual are necessary but not sufficient conditions. The sample's solid matrix composition plays also an important role. Methyl group driven relaxation via tunnelling [45] strongly influenced $^{13}C$ nuclei $T_1$ at 1 T and 4.2 K. When radical precursors molecules with methyl group were employed (i.e. TMP) the glucose relaxation time did not exceed 1 h. Getting rid of methyl group relaxation, via deuteration (i.e. dTMP) or replacement of the precursor molecule with a methyl group free one (i.e. AKG), extended the glucose relaxation time by 1 order of magnitude. Replacing the water:glycerol solvent with its deuterated counterpart further improved the $T_1$ value by a factor of two. Carbon polarization can leak through the surrounding nuclei via spin diffusion, which is still active even in absence of radicals' driven paramagnetic relaxation [46,47]. After thermalization, the abundant nuclear species in our samples (i.e. $^{1}H$ and $^{2}H$) are completely depolarized because of fast spin diffusion during heating and can act as a polarization sink for $^{13}C$ nuclei. Nevertheless, since $\gamma(^{2}H) < \gamma(^{1}H)$, spin diffusion-assisted leakage of $^{13}C$ polarization is less efficient for deuterated glassing matrices [48].

The presence of $^{1}H$ and $^{2}H$ nuclei at thermal equilibrium in the sample had another important consequence: it made the presence of permanent magnets along the extraction path a key requirement for a successful experiment. Exposing the thermalized samples to at least 100 mT allowed us to avoid spectral overlap of the $^{13}C$ NMR line with the $^{1}H$ or $^{2}H$ ones. A spectral overlap would lead to energy exchange between the different nuclear pools draining polarization from carbon nuclei by means of the mechanism known as low-field Thermal Mixing [29].

Because of dTMP limited availability and unknown biocompatibility, our radical precursor of choice for the *in-vivo* experiments became AKG, that was introduced by Gaunt et al. as a universal and high throughput radical precursor [49]. Nevertheless, at same UV-irradiation conditions, we measured a quantum yield of only 3% instead of 30% earlier reported [49]. The latter was surely due to the absence of lactic acid in our sample preparation that can act as electron acceptor during the photo-excitation process [50]. At the same time, working with a UV-radical showing a more complex spectrum with respect to the single line of TMP and dTMP, led us to interesting observation during the annihilation process. Figure S1 shows that the UV-irradiated AKG spectrum changed appearance around 175 K, when the quenching process starts: first it became narrower and then lost one 1/2 nuclear spin coupling evolving from a quadruplet to a triplet. Around 190 K, the radical concentration dropped by half, and at 200 K the radical spectrum showed a single peak to then disappear above 225 K. This evolution of the signal suggested that the radical quench process either happens in more steps, or more than one species are generated in the first place. It is important to notice that the signal evolution was not reversible: cooling back down the sample did not lead to the recovery of any feature in the spectral appearance. We are currently running more experiment to quantify this phenomenon.

Although DNP performances were similar to what we obtained earlier using TMP and dTMP [32], FM played an even more crucial role to achieve high polarization in the solid state. The reason for that was the surprisingly slow electron spins dynamics upon microwave excitation (see Figure 1D). Without FM, excitation was two times slower than relaxation. The consequence was poor spectral diffusion leading to only a small portion of the radical ESR spectrum to participate to the DNP mechanism [51,52]. Since the characteristic time constant $\tau_D$ for the electron spins excitation to spread across the entire ESR line depends on the radical spectral width $D$ and the diffusion coefficient $\Delta(\omega)$ as $\tau_D = D^2/4\Delta(\omega)$ [52], FM artificially decreased $D$, making spectral diffusion faster with respect to $T_{eR}$ that remained constant instead (see Supporting Information Figure S9). DNP improved until the excitation time constant reached a plateau for 50 MHz of FM. Higher values of FM led to a slight performance decrease of the DNP, because electron spins contributing the enhancement with opposite polarity started to be involved.

A more quantitative analysis of this phenomenon is beyond the scope of this paper and will be the object of a future publication. Nevertheless, as a comparison, we report in Supporting Information (Figure S10) the LOD-ESR and DNP investigation of the properties of a sample that does not benefit from FM: 2 M glucose dissolved in glycerol:water 1:1 (v/v) doped with 20 mM of AH111501 trityl radical. The main difference with respect to the AKG-glucose sample was the 5-times faster excitation dynamics with respect to relaxation, already for monochromatic microwaves irradiation.

In conclusion, we demonstrate that under certain circumstances hyperpolarization via DNP can be transported, and the high spin order preserved well enough to conduct *in-vivo* experiments with a contrast very similar to what ca be obtained by means of an on-site polarizer.

It is dutiful to discuss the limitations of this technique, thus to mention the case of [1-$^{13}$C]pyruvic acid. Despite high initial polarization [39], robustness and ease of operation of our CFP based methodology, a thermalization step > 10 s always depleted at least 80% of the hyperpolarized [1-$^{13}$C]pyruvic acid signal. Moreover, the remaining spin order did not survive manual extraction. These two observations were the consequence of fast relaxation arising from the presence of methyl groups in the vicinity of $^{13}$C. Using [d$_4$, 1-$^{13}$C]pyruvic acid instead, and mixing it in equal volume to deuterate glycerol:water mitigated the problem. Losses like the ones observed on the glucose sample were measured during thermalization and extraction. Nevertheless, even at 4.2 K, the carbon $T_1$ at storage conditions did not exceed 90 min (see Figure S12). Thus, HP MR contrast agents with methyl groups in their molecular structure might benefit from a more advanced and milliseconds-long thermalization and extraction procedure. Moreover, it was earlier demonstrate that changing the pyruvic acid's glass properties by means of annealing [29], can prologue its carbon half-life by one order of magnitude.

## Methods

*Samples preparation, radical generation and quench temperature*

All chemicals were purchased from Sigma-Aldrich (Brøndby, Denmark) excepted for the radical precursor deuterated trimethylpyruvic acid (dTMP) that was synthesized in house. The substrate (2 M of [U-$^{13}$C, $d_7$]-D-glucose or 5 M of HP001(bis-1,1-(hydroxymethyl)-[1-$^{13}$C]cyclopropane-$d_8$)) and the radical precursor (0.7 M of TMP or 0.7 M of dTMP or 0.7 M of AKG) were dissolved in the solvent solution (glycerol:water 1:1 (v/v) or $d_8$-glycerol:$D_2O$ 1:1 (v/v)) to obtain a final sample volume of 150 ± 5 µL; the latter was sonicated at 40 °C for 5 min to efficiently degas the sample and improve the glass quality after freezing. 15 x 10.0 ± 0.5 µL droplets were poured in liquid nitrogen to form a frozen bead. The frozen sample was transferred to a quartz Dewar (Magnettech, Berlin, Germany) filled with liquid nitrogen for UV irradiation. The irradiation set up was extensively described earlier [39]. UV-light was shined in steps on the sample for a total of 300 s and using two broad-band source (Dymax BlueWave 75, Connecticut, USA) at full power (i.e., 35 W/cm2). Radical concentration was measured immediately after each irradiation step by inserting the tail of the quartz Dewar in the cavity of an X-band spectrometer (Miniscope MS 5000, Magnettech, Berlin, Germany) and following methods described earlier [39].

To quantify the radical quench temperature of the sample, the X-band ESR spectrometer was equipped with a Variable Temperature Insert (VTI) and 1 UV-irradiated sample's bead transferred into it. The radical concentration profile as a function of temperature was characterized from 100 K to 275 K in steps of 5 K. To consider the Boltzmann factor and relate the data points' value to the number of spins only, the signal intensity was multiplied by the temperature at which it was acquired.

The glucose and HP001 samples prepared with 20 mM AH111501 trityl radicals were not UV-irradiated.

*Custom Fluid Path (CFP) design*

The CFP is built using two concentric tubes (an outer lumen - black - and an inner lumen - red - ending with a nozzle - green-) that allow gas or liquid to flow from a dispenser onto the sample and out of the sealed environment (all design details and operation's principles were reported earlier [34]). A pierced half cylinder (black cut view in the figure) is used to avoid obstruction of the nozzle's output from direct contact with the sample that sits in the vial (transparent component in the figure). For this study, the CFP was equipped with a Cernox temperature sensor (CX-1030, Lake Shore Cryotronics, Westerville, USA) to monitor sample's temperature during thermalization and extraction.

*Solid-state DNP and LOD-ESR*

All DNP and LOD-ESR were performed on a homebuilt dDNP polarizer (6.7 T magnet from Magnex, Oxford, UK) working at 1.15±0.05 K conceptually similar to the idea introduced in 2003 [18], and equipped with a 94 GHz solid-state source (Polarize Aps, Copenhagen, Denmark) coupled to a 200×2R4 frequency doubler (VDI, Charlottesville, VA, USA), which provided an output power of 100 mW at 188 GHz. The source, digitally controlled through NI-DAQ device USB-6525 (National Instruments, Austin, TX, USA) had a tuning range of ±1 GHz and the possibility to modulate the output frequency at a rate up to 10 kHz and with an amplitude of up to 500 MHz. All $^{13}C$ NMR acquisitions were performed using a Varian INOVA console (Palo Alto, CA, USA) connected to a low temperature probe equipped with permanent magnets along the central loading tube [34], placed inside the polarizer VTI (Variable Temperature Insert), remotely tuned to 71.8 MHz.

For LOD-ESR measurements a home-built spectrometer was used [39,40]. For radical spins relaxation and saturation time constant measurements the output power of the microwave source was modulated at 0.5 Hz between 0 mW and 50 mW. The rate was low enough to record the full-time evolution of the electron spins during saturation and relaxation; the signal was averaged 256 times. Extraction of the relaxation and excitation time constants ($T_c$) were performed by fitting the signal evolution in the time domain to the equation $S(t) = A\left(exp(-t/T_c) - exp(-t/\tau)\right)$ where $A$ is a proportionality factor depending on the sample properties and measuring parameters and $\tau$ the time constant of the measuring split solenoid coil. By measuring the time constant $\tau$ exciting the coil through a step function, the electron $T_c$ was the only free parameter of the fit. For radical spectrum recording, the microwave frequency was increased from 187.8 GHz to 188.3 GHz in steps of 1.25 MHz and the output power modulated at 4.8 Hz between 0 mW and 50 mW, and this signal fed into a lock-in amplifier. For each frequency step the demodulated signal was integrated for 20 s in the time domain, equivalent to set the low pass filter of the lock-in to 0.05 Hz.

*Thermalization, extraction and NMR measurements from the transportable cryostat*

To thermalize the sample and quench the radicals, the CFP was raised by 10 cm pulling the sample out of the NMR detection coil and positioned above the liquid He level. At this point, microwaves were switched off, the top part of the CFP was connected to a room temperature He gas dispenser, equipped with a pneumatic valve, and whose flow was controlled by a feedback loop on the sample's temperature. He gas was blown continuously for approximately 4 s at 10 bar until the radicals' quench temperature of 225 K was achieved, and then blown intermittently for 6 s more to thermalize the full sample's volume while avoiding extra heating and possible polarization losses. Subsequently, the CFP was lowered inside the coil to cool down the sample and measure NMR.

After thermalization, the sample was prepared for extraction. The portable cryostat, equipped with a magnetically enforced probe[38], was precooled to 4.2 K and placed at the bottom of the dDNP polarizer. The extraction procedure happened in 3 steps: firstly, the CFP was pulled up and the sample's vial raised to the dDNP polarizer's loading chamber without breaking the vacuum; secondly, the polarizer's gate valve was closed and the loading chamber disconnected; thirdly, the loading chamber was manually docked at the top of the portable cryostat, its gate valve opened and the sample lowered inside the Halbach magnet. At this point NMR could be performed on the extracted sample to evaluate its relaxation properties in storage conditions. The solenoidal coil placed inside the Halbach magnet was remotely tuned and matched outside of the cryostat by means of a T/M box with two piston trimmer capacitors (Voltronics V1949). The Halbach magnet has a $^{13}C$ resonance frequency of 10.0 MHz and 10.2 MHz at 77 K and 4.2 K, respectively. All NMR $T_1$ measurements were performed using a compact bench-top spectrometer (Kea2, Magritek, Wellington, New Zealand) by applying 5° rf pulses every 5 min at 4.2 K and every 1 min at 77 K. The pulse angle was previously calibrated on a sample sending a train of 250 pulses spaced by 10 ms and with a length of 10 μs and a power of 5 W made the signal to decay. The decay was fit with the equation $S(n) = S(0)cos(\theta)^{n-1}$, where $\theta$ is the flip angle and $n$ the number of acquisitions.

*Animals handling*

All animal experiments were approved by the Danish Animal Inspectorate. Female Wistar rats (8 weeks old, weight 200g) from Taconic Biosciences (Denmark) were included. The rats were anaesthetized with 2.5-3 % sevoflurane in 2 l/min medical air. A tail vein catheter was placed for infusion of hyperpolarized glucose, and normothermia was maintained using an MRI compatible small-animal monitoring system (Small Animal Instruments Inc, USA). The animals underwent imaging of the kidneys.

*Magnetic resonance imaging with hyperpolarized Glucose*

MRI was performed on a 3T scanner (MR750, GE Healthcare, USA) with a $^{13}C/^{1}H$ rat volume coil (RAPID Biomedical, Germany). A volume of 1 ml was injected through a tail vein catheter. Anatomical images were acquired for reference. A fast spin echo sequence was used for the body (1500 ms repetition time, 11 ms echo time, 24 echo train length, 4 mm slice thickness, 160 × 160 matrix for a 160 x 160 mm field-of-view, flip angle = 16°). Hereafter, $^{13}C$ CSI was performed and images were acquired (65 ms repetition time, 10 x 10 matrix for a 120 x 120 mm field-of-view, spectral resolution = 1024 Hz, bandwidth = 20.000 Hz, flip angle = 10°). The transmit gain was calibrated using a phantom with appropriate load and kept constant throughout the experiment. The carbon center frequency was extrapolated from the proton frequency and kept constant within the same animal.

*Magnetic resonance imaging with hyperpolarized HP001*

A 2D gradient echo spiral sequence was designed with the following parameter: FOV:120 x 120 mm, resolution = 1x1mm, readout time = 21ms, matrix size = 120 x 120. In-plane center-out 10 arms spiral trajectory was designed as variable density to minimize the ringing artefact.

All imaging experiments were performed on GE 3T scanner with $^{1}H/^{13}C$ dual turned rat coil. $^{13}C$ dynamic angiogram images are acquired as 2D coronal view projection. The acquisition started after starting the injection. The imaging parameters were: FOV = 120 x 120 mm, resolution = 1 x 1 mm, TR/TE = 44/1 ms, flip angle = 5º, image frame time= 1/0.5 s, number of frames = 40. Only the first 16 frames are shown in the manuscript.

*Magnetic field measurements and simulations*

To plot the magnetic field on the bottom panel of Figure 3, magnetic field simulations were performed using MATLAB (MathWorks, Natick, MA, USA) and COMSOL 5.4 (COMSOL Multiphysics, Burlington, Massachusetts, USA) [34,38]. Moreover, the magnetic field along the extraction path was measured with a Hall probe Lake Shore 475 (Lake Shore Cryotronics, Westerville, OH, USA) to verify the accuracy of the simulations. The temperature profile as a function of time was used to convert the magnetic field vs distance plot to magnetic field vs time plot.

*Processing and data analysis*

All data were processed and analyzed in MATLAB (MathWorks, Natick, MA, USA). All numerical results are reported in the main text as average of repeated measurements, and the standard deviation represents the error. All measurements were repeated at least three times.

## References

1. Garus-Pakowska, A. Metabolic Diseases—A Challenge for Public Health in the 21st Century. *Int. J. Environ. Res. Public. Health* **20**, 6789 (2023).
2. van de Weijer, T. & Schrauwen-Hinderling, V. B. Application of Magnetic Resonance Spectroscopy in metabolic research. *Biochim. Biophys. Acta BBA - Mol. Basis Dis.* **1865**, 741–748 (2019).
3. Haris, M. *et al.* Molecular magnetic resonance imaging in cancer. *J. Transl. Med.* **13**, 313 (2015).

4. Zilberter, Y. & Zilberter, M. The vicious circle of hypometabolism in neurodegenerative diseases: Ways and mechanisms of metabolic correction: Hypometabolism in Neurodegenerative Diseases. *J. Neurosci. Res.* **95**, 2217–2235 (2017).
5. Cunningham, C. H. *et al.* Hyperpolarized$^{13}$ C Metabolic MRI of the Human Heart: Initial Experience. *Circ. Res.* **119**, 1177–1182 (2016).
6. Kelloff, G. J. *et al.* Progress and Promise of FDG-PET Imaging for Cancer Patient Management and Oncologic Drug Development. *Clin. Cancer Res.* **11**, 2785–2808 (2005).
7. Long, N. M. & Smith, C. S. Causes and imaging features of false positives and false negatives on 18F-PET/CT in oncologic imaging. *Insights Imaging* **2**, 679–698 (2011).
8. Patel, S. *et al.* Simultaneous noninvasive quantification of redox and downstream glycolytic fluxes reveals compartmentalized brain metabolism. *Sci. Adv.* **10**, eadr2058 (2024).
9. Gallagher, F. A. *et al.* Hyperpolarized$^{13}$ C MRI and PET: In Vivo Tumor Biochemistry. *J. Nucl. Med.* **52**, 1333–1336 (2011).
10. Golman, K., Thaning, M., & others. Real-time metabolic imaging. *Proc. Natl. Acad. Sci.* **103**, 11270–11275 (2006).
11. Golman, K. *et al.* Parahydrogen-induced polarization in imaging: Subsecond $^{13\$}$C angiography. *Magn. Reson. Med.* **46**, 1–5 (2001).
12. Comment, A. & Merritt, M. E. Hyperpolarized Magnetic Resonance as a Sensitive Detector of Metabolic Function. *Biochemistry* **53**, 7333–7357 (Dec 2).
13. Brindle, K. M. Imaging Metabolism with Hyperpolarized C-13-Labeled Cell Substrates. *J. Am. Chem. Soc.* **137**, 6418–6427 (2015).
14. Hesketh, R. L. & Brindle, K. M. Magnetic resonance imaging of cancer metabolism with hyperpolarized 13C-labeled cell metabolites. *Curr. Opin. Chem. Biol.* **45**, 187–194 (2018).

15. Reineri, F., Boi, T. & Aime, S. ParaHydrogen Induced Polarization of 13C carboxylate resonance in acetate and pyruvate. *Nat. Commun.* **6**, 5858 (2015).

16. Cavallari, E. *et al.* The 13C hyperpolarized pyruvate generated by ParaHydrogen detects the response of the heart to altered metabolism in real time. *Sci. Rep.* **8**, 8366 (2018).

17. De Maissin, H. *et al.* In Vivo Metabolic Imaging of [1-$^{13}$ C]Pyruvate-$d_3$ Hyperpolarized By Reversible Exchange With Parahydrogen**. *Angew. Chem. Int. Ed.* **62**, e202306654 (2023).

18. Ardenkjaer-Larsen, J. H. *et al.* Increase in signal-to-noise ratio of > 10,000 times in liquid-state NMR. *Proc. Natl. Acad. Sci. U. S. A.* **100**, 10158–10163 (Sep 2).

19. Wodtke, P., Grashei, M. & Schilling, F. Quo Vadis Hyperpolarized 13C MRI? *Z. Für Med. Phys.* S0939388923001204 (2023) doi:10.1016/j.zemedi.2023.10.004.

20. Yoshihara, H. A. I. *et al.* High-field dissolution dynamic nuclear polarization of [1-C-13]pyruvic acid. *Phys. Chem. Chem. Phys.* **18**, 12409–12413 (2016).

21. Ardenkjær-Larsen, J. H. *et al.* Cryogen-Free dissolution Dynamic Nuclear Polarization polarizer operating at 3.35 T, 6.70 T and 10.1 T. *Magn. Reson. Med.* **accepted**, (2018).

22. Capozzi, A. *et al.* Gadolinium Effect at High-Magnetic-Field DNP: 70% $^{13}$ C Polarization of [U- $^{13}$ C] Glucose Using Trityl. *J. Phys. Chem. Lett.* **10**, 3420–3425 (2019).

23. Ardenkjaer-Larsen, J. H. *et al.* Dynamic nuclear polarization polarizer for sterile use intent. *NMR Biomed.* **24**, 927–932 (2011).

24. Nelson, S. J. *et al.* Metabolic Imaging of Patients with Prostate Cancer Using Hyperpolarized [1-C-13]Pyruvate. *Sci. Transl. Med.* **5**, 198ra108 1–10 (2013).

25. Gallagher, F. A. *et al.* Imaging breast cancer using hyperpolarized carbon-13 MRI. *Proc. Natl. Acad. Sci.* **117**, 2092–2098 (2020).

26. Qin, H. *et al.* Clinical translation of hyperpolarized $^{13}$ C pyruvate and urea MRI for simultaneous metabolic and perfusion imaging. *Magn. Reson. Med.* **87**, 138–149 (2022).

27. Ardenkjaer-Larsen, J. H. On the present and future of dissolution-DNP. *J. Magn. Reson.* **264**, 3–12 (2016).

28. Ardenkaer-Larsen, J. H., Axelsson, O. H. E., Golman, K. K., Wolber, J. & Howard, M. Methods and devices for hyperpolarising and melting NMR samples in a cryostat. *US Pat. Trademark Off. Electron. Off. Gaz. Pat.* (2006).

29. Hirsch, M. L., Kalechofsky, N., Belzer, A., Rosay, M. & Kempf, J. G. Brute-Force Hyperpolarization for NMR and MRI. *J. Am. Chem. Soc.* **137**, 8428–8434 (2015).

30. Hirsch, M. L. *et al.* Transport and imaging of brute-force 13 C hyperpolarization. *J. Magn. Reson.* **261**, 87–94 (2015).

31. Blumberg, W. E. Nuclear Spin-Lattice Relaxation Caused by Paramagnetic Impurities. *Phys. Rev.* **119**, 79–84 (1960).

32. Capozzi, A. *et al.* Efficient Hyperpolarization of U-(13) C-Glucose Using Narrow-Line UV-Generated Labile Free Radicals. *Angew Chem Int Ed Engl* **58**, 1334–1339 (2019).

33. Capozzi, A., Cheng, T., Boero, G., Roussel, C. & Comment, A. Thermal annihilation of photo-induced radicals following dynamic nuclear polarization to produce transportable frozen hyperpolarized 13C-substrates. *Nat. Commun.* **8**, 15757 (2017).

34. Capozzi, A. *et al.* Metabolic contrast agents produced from transported solid 13C-glucose hyperpolarized via dynamic nuclear polarization. *Commun. Chem.* **4**, 95 (2021).

35. Eichhorn, T. R. *et al.* Hyperpolarization without persistent radicals for in vivo real-time metabolic imaging. *Proc. Natl. Acad. Sci. U. S. A.* **110**, 18064–18069 (Nov 5).

36. Ji, X. *et al.* Transportable hyperpolarized metabolites. *Nat. Commun.* **8**, (2017).

37. El Daraï, T. *et al.* Porous functionalized polymers enable generating and transporting hyperpolarized mixtures of metabolites. *Nat. Commun.* **12**, 4695 (2021).

38. Capozzi, A. Design and performance of a small bath cryostat with NMR capability for transport of hyperpolarized samples. *Sci. Rep.* **12**, 19260 (2022).

39. Capozzi, A., Karlsson, M., Petersen, J. R., Lerche, M. H. & Ardenkjaer-Larsen, J. H. Liquid-State $^{13}$C Polarization of 30% through Photoinduced Nonpersistent Radicals. *J. Phys. Chem. C* **122**, 7432–7443 (2018).

40. Lê, T. P., Hyacinthe, J.-N. & Capozzi, A. How to improve the efficiency of a traditional dissolution dynamic nuclear polarization (dDNP) apparatus: Design and performance of a fluid path compatible dDNP/LOD-ESR probe. *J. Magn. Reson.* **338**, 107197 (2022).

41. Wenckebach, W. Th. Dynamic nuclear polarization via the cross effect and thermal mixing: A. The role of triple spin flips. *J. Magn. Reson.* **299**, 124–134 (2019).

42. Wenckebach, W. Th. Dynamic nuclear polarization via the cross effect and thermal mixing: B. Energy transport. *J. Magn. Reson.* **299**, 151–167 (2019).

43. Ardenkjær-Larsen, J. H. *et al.* Cryogen-free dissolution dynamic nuclear polarization polarizer operating at 3.35 T, 6.70 T, and 10.1 T. *Magn. Reson. Med.* **81**, 2184–2194 (2019).

44. Ferrari, A. *et al.* Performance and reproducibility of 13C and 15N hyperpolarization using a cryogen-free DNP polarizer. *Sci. Rep.* **12**, 11694 (2022).

45. Latanowicz, L. NMR relaxation study of methyl groups in solids from low to high temperatures. *Concepts Magn. Reson. Part A* **27A**, 38–53 (2005).

46. Abragam, A. & Goldman, M. *Order and Disorder*. (Clarendon Press, Oxford, 1982).

47. Abragam, A. *The Principles of Nuclear Magnetism*. (University Press, Oxford, 1961).

48. Parish, C. *et al.* Effects of glassing matrix deuteration on the relaxation properties of hyperpolarized 13C spins and free radical electrons at cryogenic temperatures. *J. Chem. Phys.* **150**, 234307 (2019).

49. Gaunt, A. P. *et al.* Labile Photo-Induced Free Radical in α-Ketoglutaric Acid: a Universal Endogenous Polarizing Agent for In Vivo Hyperpolarized $^{13}$ C Magnetic Resonance. *Angew. Chem.* **134**, (2022).

50. Lewis, J. S., Gaunt, A. P. & Comment, A. Photochemistry of pyruvic acid is governed by photo-induced intermolecular electron transfer through hydrogen bonds. *Chem. Sci.* **13**, 11849–11855 (2022).

51. Kundu, K., Feintuch, A. & Vega, S. Electron–Electron Cross-Relaxation and Spectral Diffusion during Dynamic Nuclear Polarization Experiments on Solids. *J. Phys. Chem. Lett.* **9**, 1793–1802 (2018).

52. Wenckebach, W. T. Spectral diffusion and dynamic nuclear polarization: Beyond the high temperature approximation. *J. Magn. Reson.* **284**, 104–114 (2017).

## Acknowledgments

This work was supported by the Danish National Research Foundation (DNRF124); the European Union's Horizon 2020 research and innovation program ERC Synergy grant 856432 - HyperQ; and the Swiss National Fund under the SPARK grant agreement no. CRSK-2_190547 and Ambizione grant agreement no. PZ00P2_193276.

We thank Prof Olivier Ouari and Dr Saket Patel for synthesizing $d_9$-TMP UV-radical precursor, and M. Duy Anh Dang for his constant help with experiments at Aarhus University Hospital.

## Author contributions

Capozzi, Lerche, Ardenkjær-Larsen and Laustsen study conception and design. Capozzi, Karlsson, Lerche, Hansen, Bertelsen and Zhao acquisition of data. Capozzi, Hansen, Zhao, and Lerche analysis and interpretation of data. Capozzi and Lerche drafting of manuscript. Laustsen, Ardenkjær-Larsen and Karlsson critical revision.

## Competing interests

Prof. Ardenkjær-Larsen is the owner of Polarize ApS. Dr Andrea Capozzi is currently employed by Polarize ApS. Polarize ApS sells dDNP equipment for pre-clinical studies.

**Correspondence** and requests for materials should be addressed to **Dr Andrea Capozzi** (andrea.capozzi@epfl.ch).

### Statistical analysis

All numerical results are reported in the main text as average of repeated measurements, and the standard deviation represents the error. All measurements were repeated at least three times

## Data availability

The authors declare that all data supporting the findings of this study are available within the paper and its Supporting Information files. Raw data are available from the corresponding author (andrea.capozzi@epfl.ch) on reasonable request.

# SUPPORTING INFORMATION

# In-vivo real-time $^{13}$C-MRSI without polarizer on site: across cities transportable hyperpolarization using UV-induced labile radicals

Andrea Capozzi [1,2*], Magnus Karlsson [2], Yupeng Zhao [2], Jan Kilund [2], Esben Søvsø Szocska Hansen [3], Lotte Bonde Bertelsen [3], Christoffer Laustsen [3], Jan Henrik Ardenkjær-Larsen [2], and Mathilde H. Lerche [2].

*[1] LIFMET, Department of Physics, EPFL, Station 6 (Batiment CH), 1015 Lausanne (Switzerland).*

*[2] HYPERMAG, Department of Health Technology, Technical University of Denmark, Building 349, 2800 Kgs Lyngby (Denmark).*

*[3] The MR Center, Department of Clinical Medicine, Aarhus University, Palle Juul-Jensens Boulevard 99, 8200 Aarhus N (Denmark)*

**Corresponding author**

*Dr. Andrea Capozzi

EPFL SB IPHYS LIFMET

CH F0 632 (Bâtiment CH), Station 6, CH-1015 Lausanne

Email: andrea.capozzi@epfl.ch; ORCID: 0000-0002-2306-9049

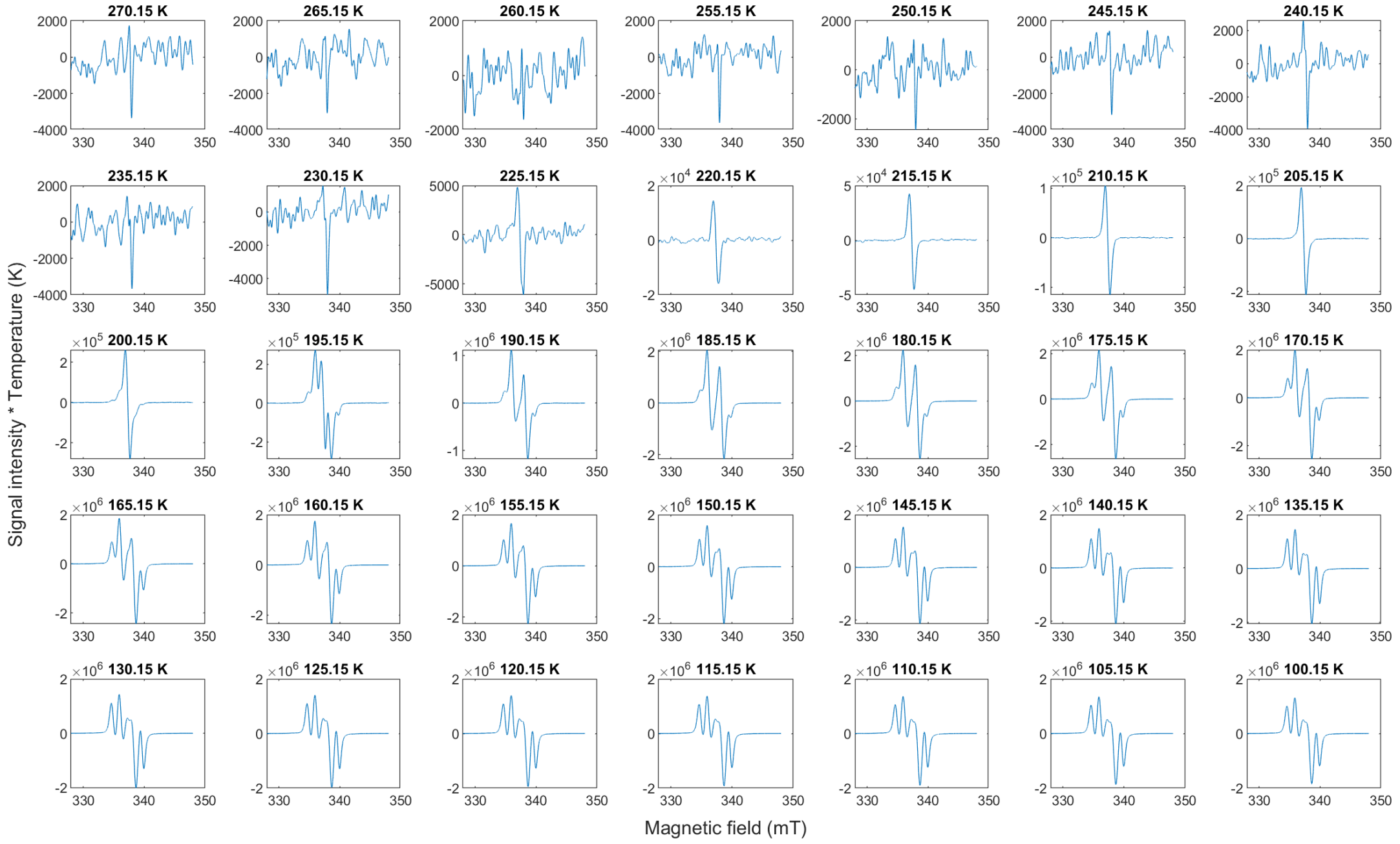

**Figure S1. Radicals quench dynamic for AKG:** one 10 µL bead of 2M glucose and 0.7 M AKG dissolved in $d_8$-glycerol:$D_2O$ 1:1 (v/v) was irradiated for 300 s vial UV-light at 35 W/cm$^2$ and transferred in the VTI of an X-band ESR spectrometer. The temperature was increased in steps of 5 K from 100.15 K to 275.15 K. At each temperature step the ESR signal was acquired. No signal could be detected above 230.15 K, thus, 225.15 K was set as heating threshold for the thermalization experiment.

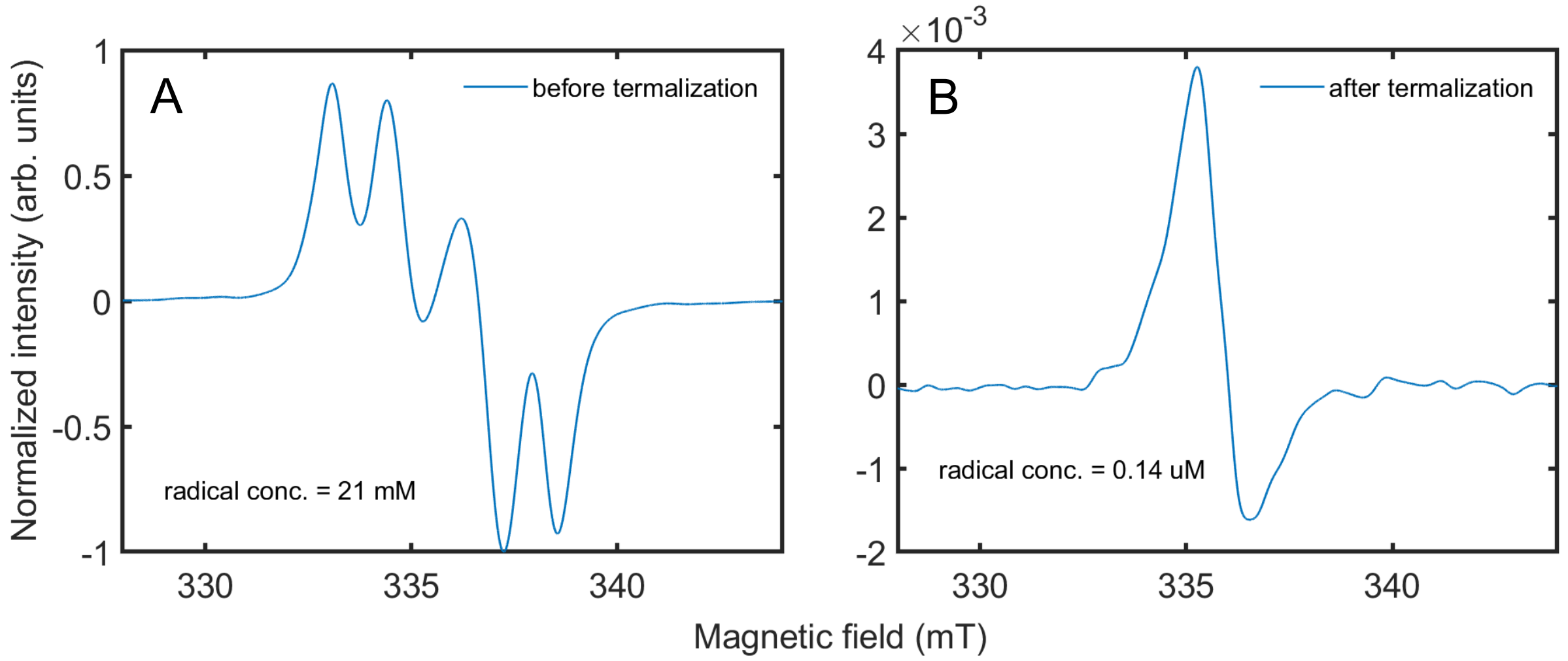

**Figure S2. Radical's left-over after thermalization:** ESR signal of the glucose sample before the DNP experiment (**A**). ESR signal of the glucose sample after thermalization (**B**); the heating process by means of He gas quenched more that 99.5% of the initial radical concentration.

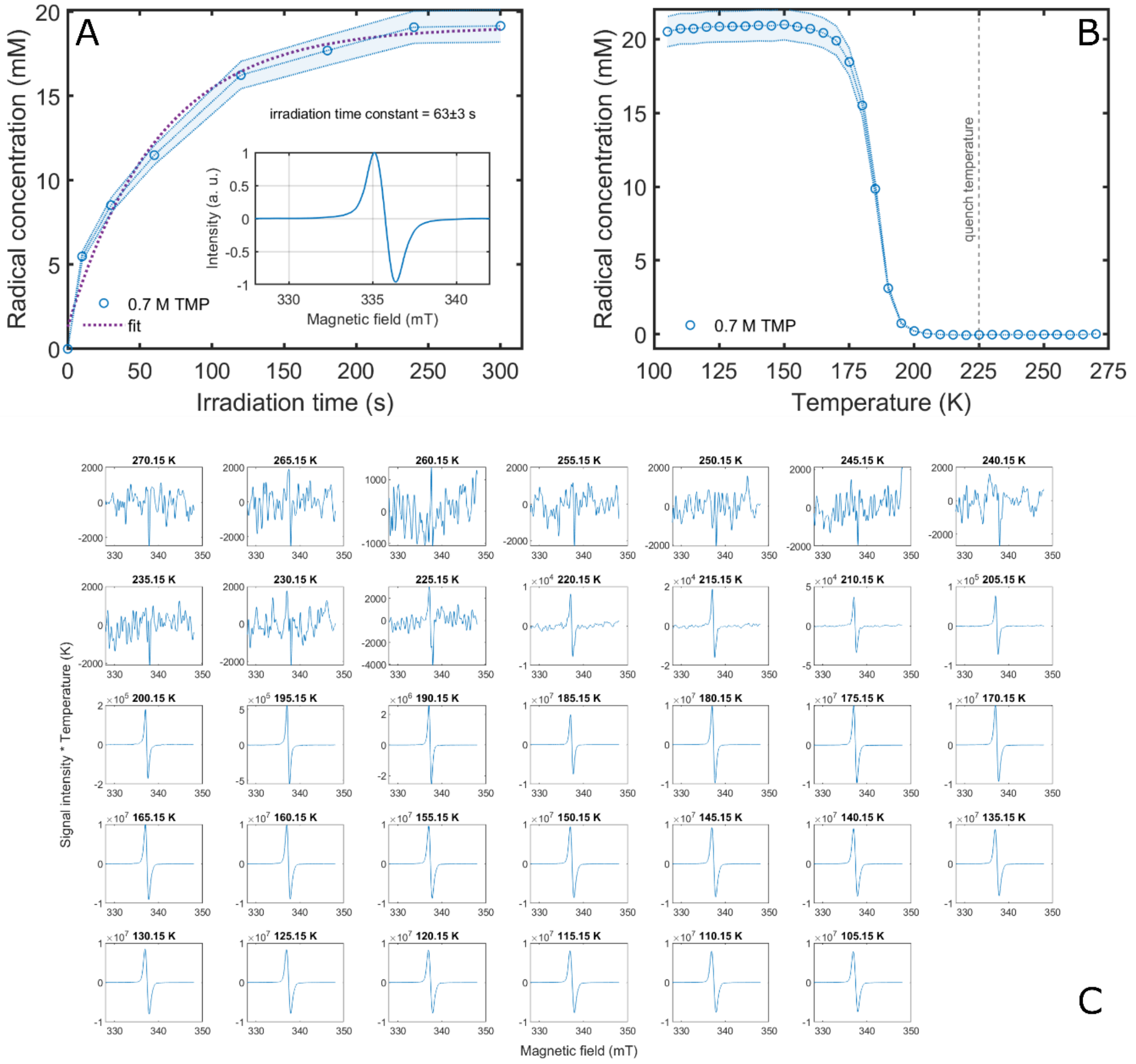

**Figure S3. Radical's properties of the glucose sample with TMP as precursor:** radical generation by means of UV-light irradiation in liquid nitrogen (**A**), quench temperature measurement (**B**), and quench temperature dynamic (**C**) for a sample containing 2M glucose and 0.7 M TMP dissolved in glycerol:water 1:1 (v/v).

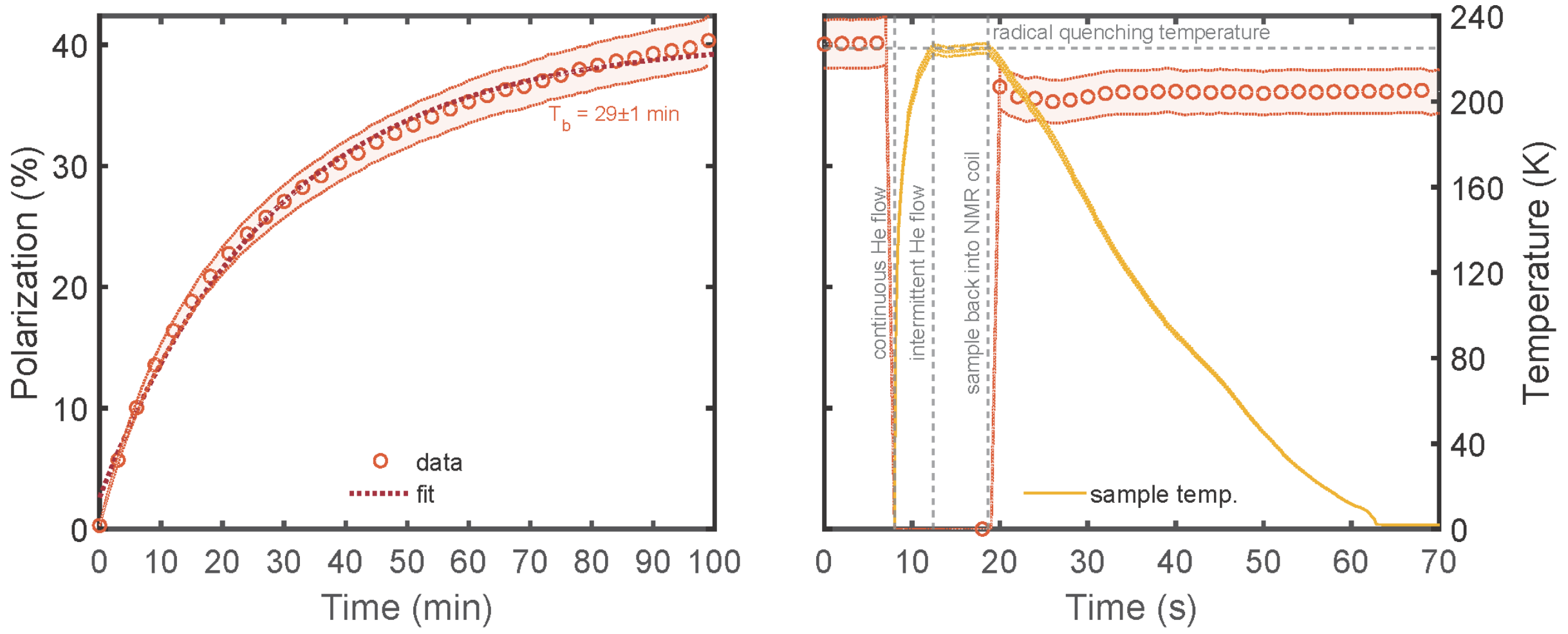

**Figure S4. Effect of solvent deuteration on DNP performance:** polarization (A) and thermalization (B) for a glucose sample prepared with a protonated solvent and AKG as radical precursor. Deuterating the solvent, increased the maximum achievable polarization by 10%; deuteration had no effect on the behavior of the sample during thermalization.

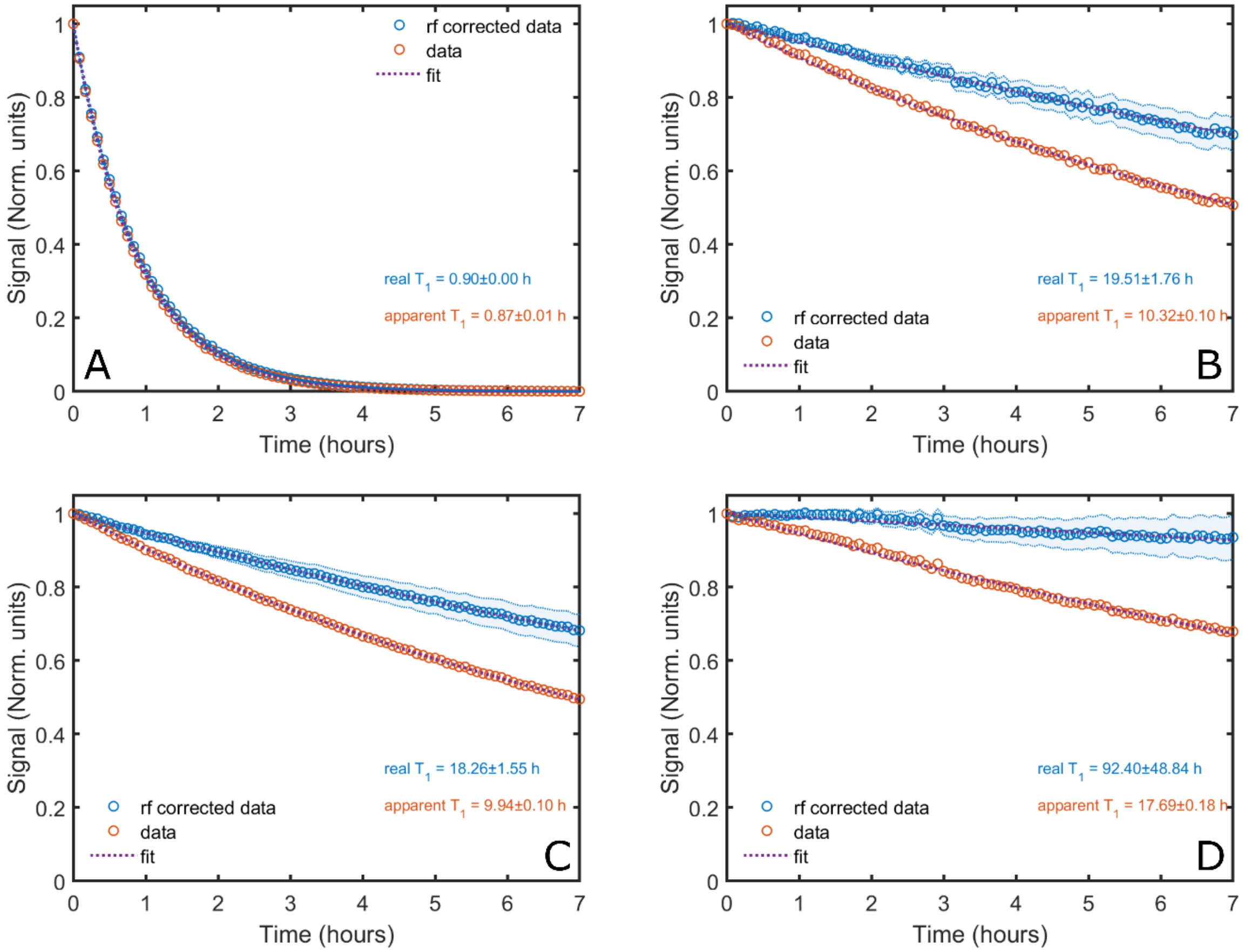

**Figure S5. Relaxation at transport conditions:** $T_1$ measurements at 4 K and 1 T by means of 5° excitation pulses every 5 min of hyperpolarized and thermalized glucose samples prepared with TMP as radical precursor and a protonated solvent (A); dTMP as radical precursor and a protonated solvent (B); AKG as radical precursor and a protonated solvent (C); AKG as radical precursor and a deuterated solvent (D). In al panels, the red circles represent the integral of the spectrum measured from the spectrometer; the blue circles were corrected for the rf pulse. The blue shaded areas represent the error on the $T_1$ values coming from the error on the estimation of the flip angle.

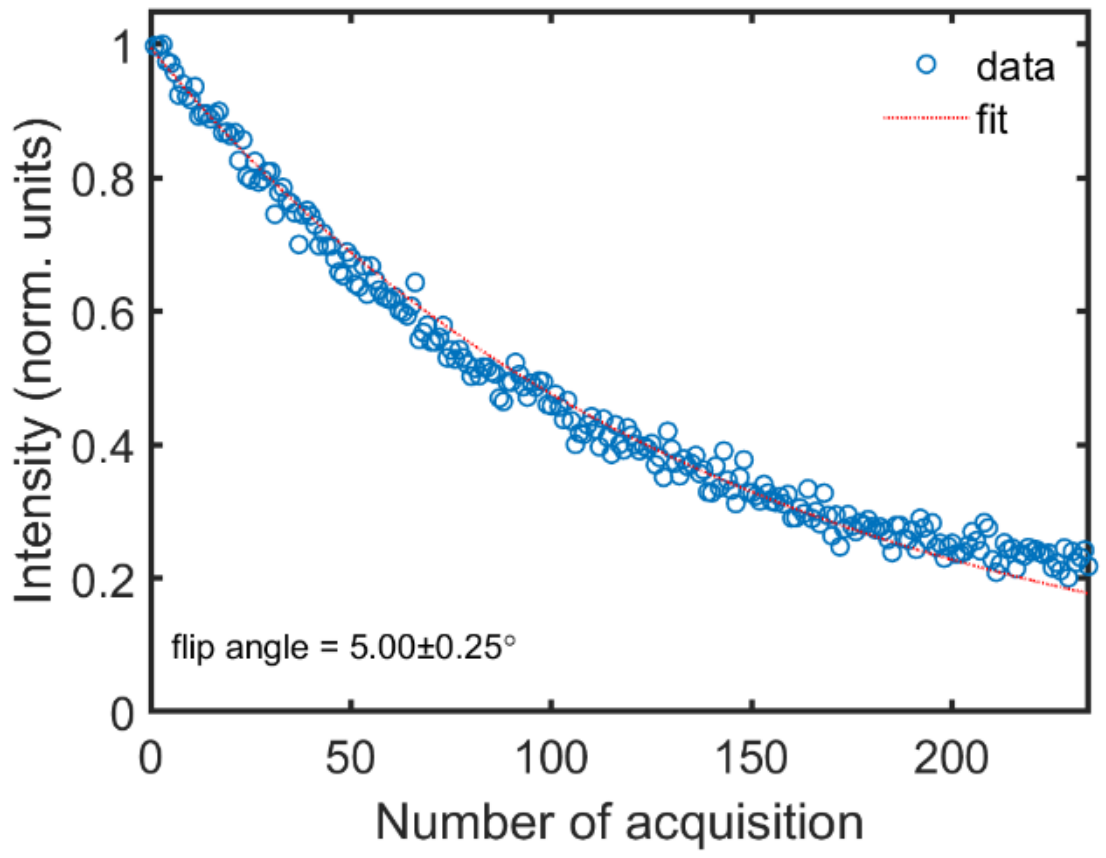

**Figure S6. Portable cryostat rf pulse calibration on $^{13}C$:** one hyperpolarized and thermalized glucose sample was used to calibrate the pulse angle of the NMR probe of the transportable cryostat. A train of 250 pulses spaced by 10 ms and with a length of 10 µs and a power of 5 W made the signal to decay. The decay was fit with the equation $S(n) = S(0)cos(\theta)^{n-1}$, where $\theta$ is the flip angle and $n$ the number of acquisitions.

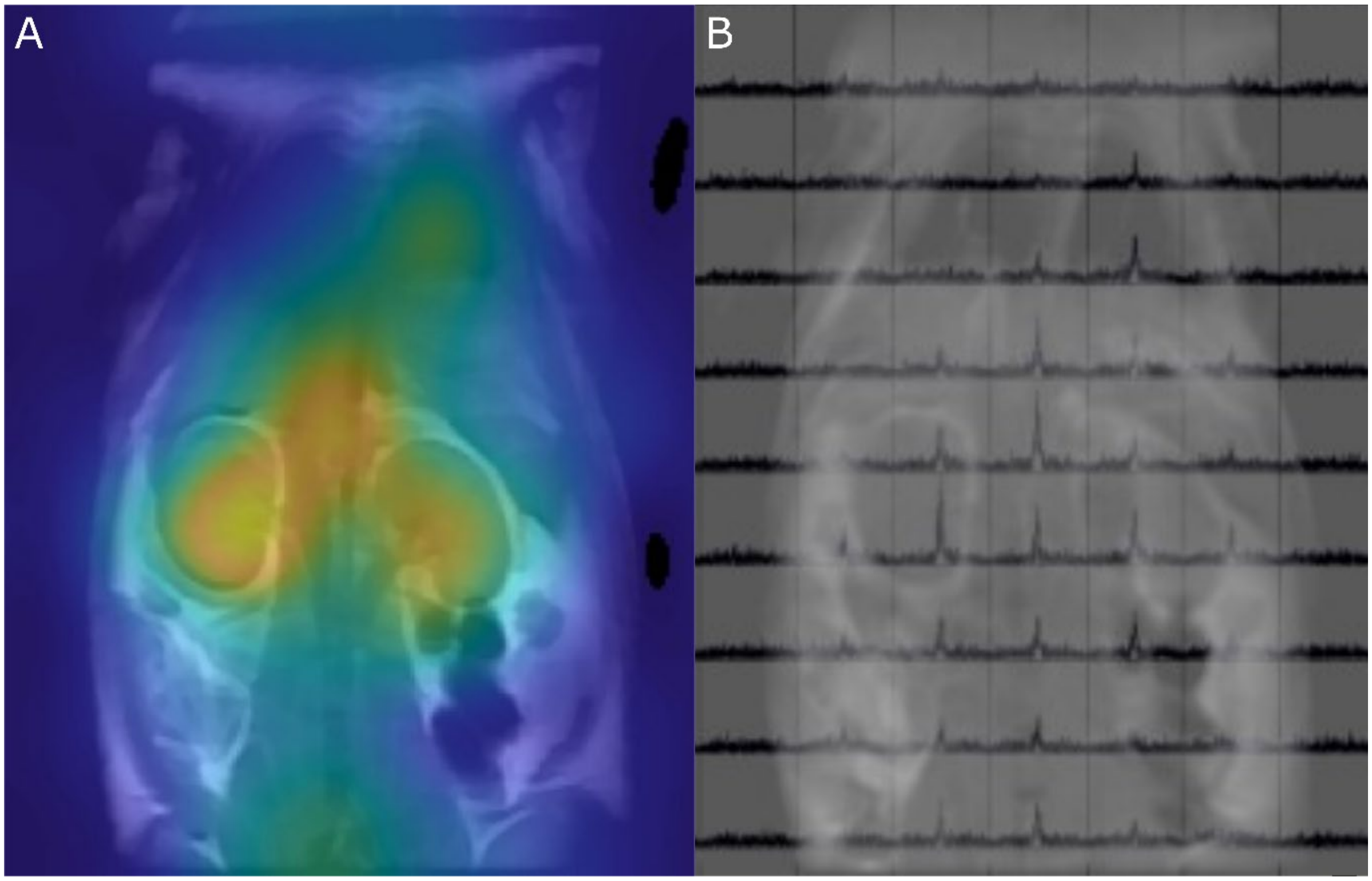

**Figure S7. Chemical Shift Imaging (CSI) of glucose in a healthy rat model:** MRI was performed on a 3T scanner (MR750, GE Healthcare, USA) with a $^{13}C/^{1}H$ rat volume coil (RAPID Biomedical, Germany). A volume of 1 ml was injected through a tail vein catheter. Anatomical images were acquired for reference. A fast spin echo sequence was used for the body (1500 ms repetition time, 11 ms echo time, 24 echo train length, 4 mm slice thickness, 160 × 160 matrix for an 160 x 160 mm field-of-view, flip angle = 16°). Hereafter, $^{13}C$ CSI was performed and images were acquired (65 ms repetition time, 10 x 10 matrix for a 120 x 120 mm field-of-view, spectral resolution = 1024 Hz, bandwidth = 20.000 Hz, flip angle = 10°). The transmit gain was calibrated using a phantom with appropriate load and kept constant throughout the experiment. The carbon center frequency was extrapolated from the proton frequency and kept constant within the same animal.

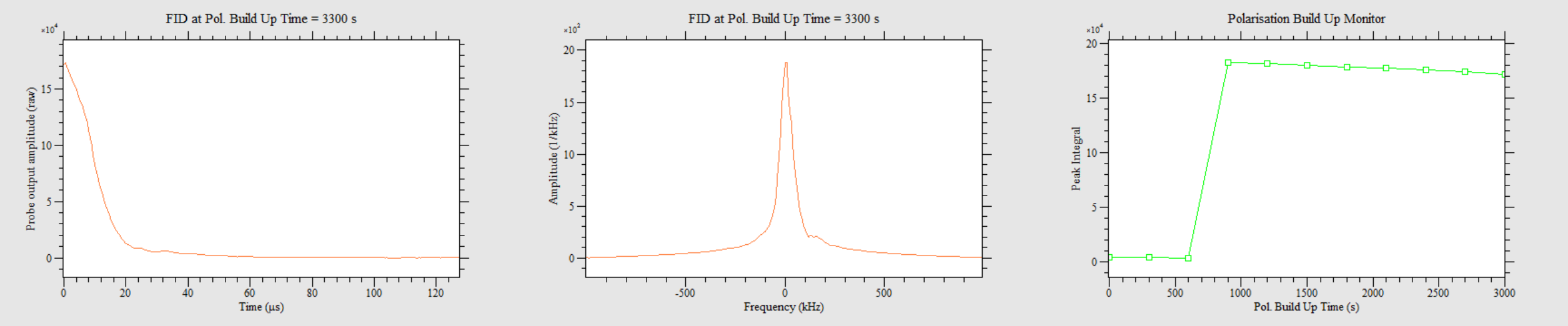

**Figure S8. Signal check after extraction of the HP001 sample (spectrometer print screen):** before attempting transport to Aarhus University Hospital, the signal of the extracted sample was first checked and compared to previous experiments. We report here a print screen of the spectrometer before disconnecting the electronics of the transportable cryostat. The three panel, from left to right, represent the signal FID, its Fourir Transform (NMR spectrum), and the integral of the spectrum as a function of time. Points were acquired every 5 min. The acquisition was started before insertion of the sample into the cryostat to get a base line reference and evaluate the decay. Before departure, the signal was observed for 45 min and negligible signal loss detected. During transport, no NMR pulsing was applied to conserve as much signal as possible.

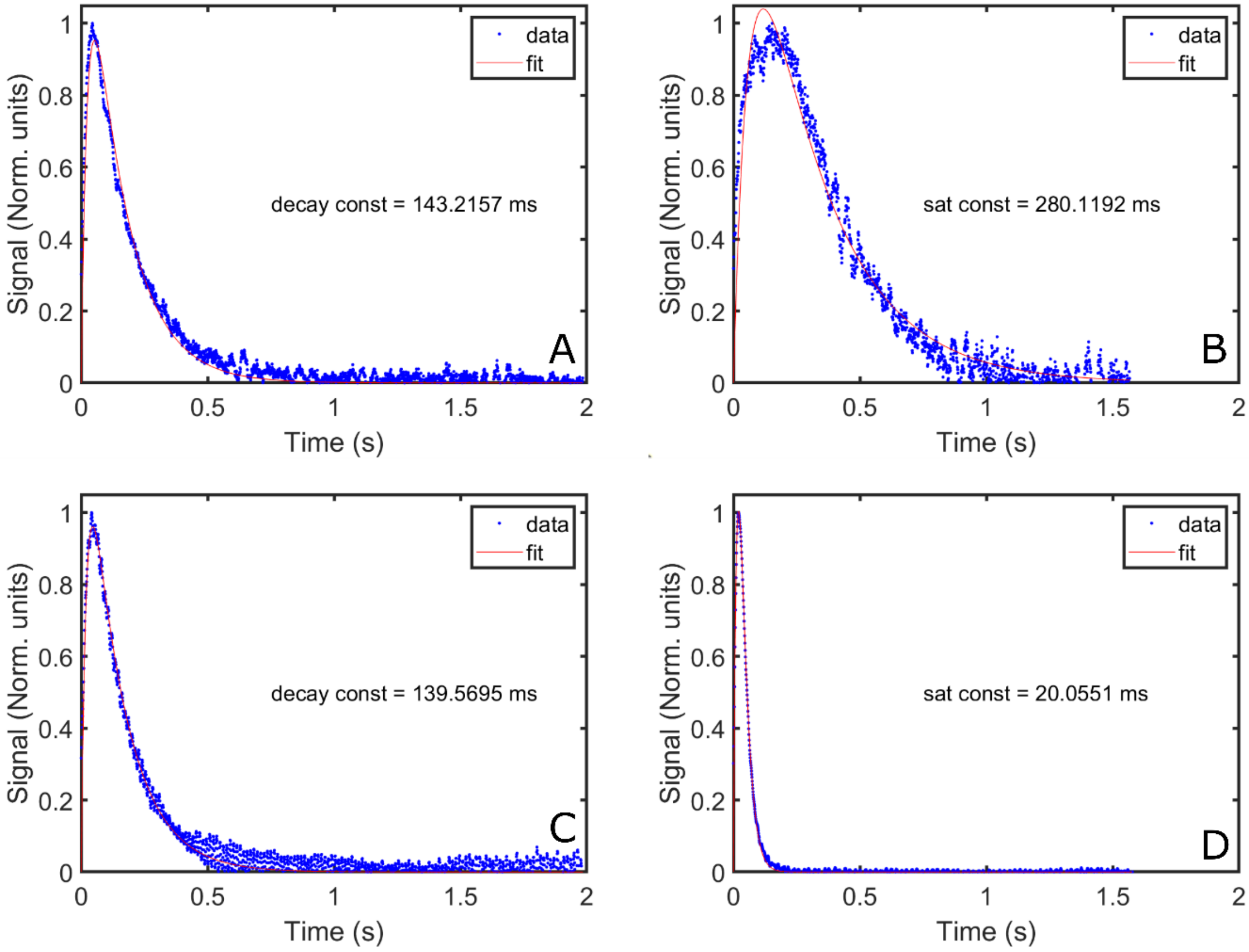

**Figure S9. Effect of FM on electron excitation and relaxation:** electron spins dynamic, measured by means of LOD-ESR [1], of the radicals induced on a sample containing 2 M glucose and 0.7 M AKG in a deuterated solvent upon switching OFF of the microwaves after excitation with no FM (**A**); switching ON of the microwaves after relaxation with no FM (**B**); switching OFF of the microwaves after excitation with 50 MHz FM (**A**); switching ON of the microwaves after relaxation with 50 MHz FM (**D**). While relaxation was FM insensitive, excitation became much faster when FM was applied.

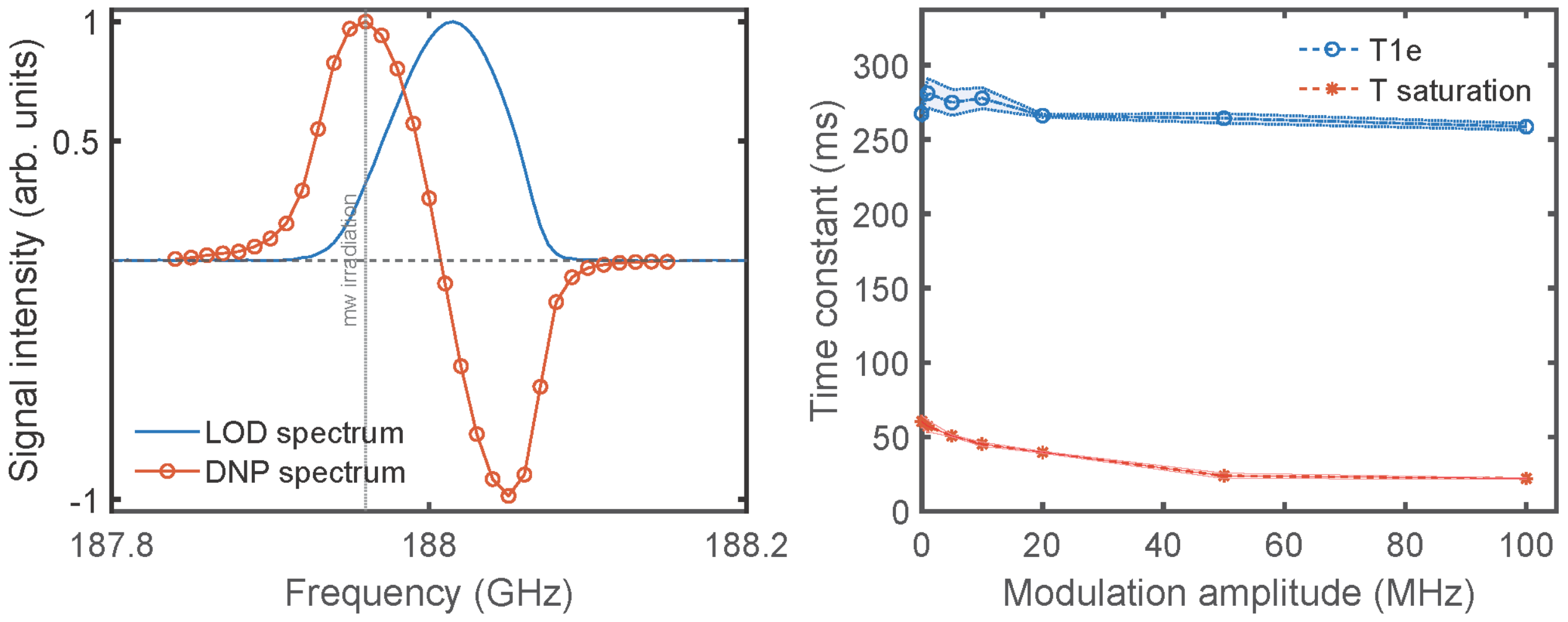

**Figure S10. LOD-ESR and DNP characterization of a 2M glucose sample doped with 20 mM of trityl radical:** as a comparison we report the LOD-ESR and DNP investigation of a sample that did not benefit from FM (i.e. no increase in DNP enhancement). The ESR spectrum and DNP spectrum are overlapped in panel **A**. The relaxation and excitation dynamics of the radicals spins as a function of FM is reported in panel **B**. It is important noticing that, differently from the AKG sample, excitation was much faster than relaxation already for monochromatic microwave irradiation.

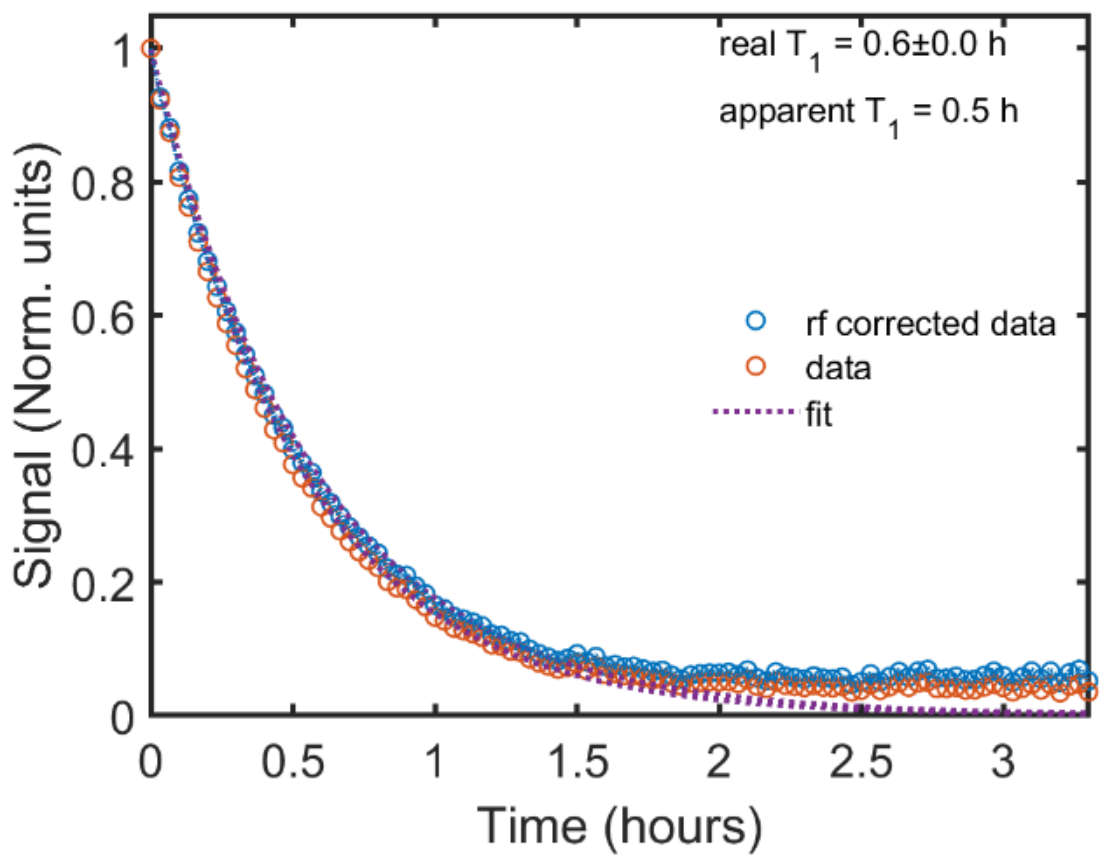

**Figure S11. Relaxation 1 T and 77 K:** The relaxation of the sample prepared with 2M glucose and 0.7 AKG dissolved in a deuterated solvent is reported after thermalization from the transportable cryostat filled with liquid nitrogen.

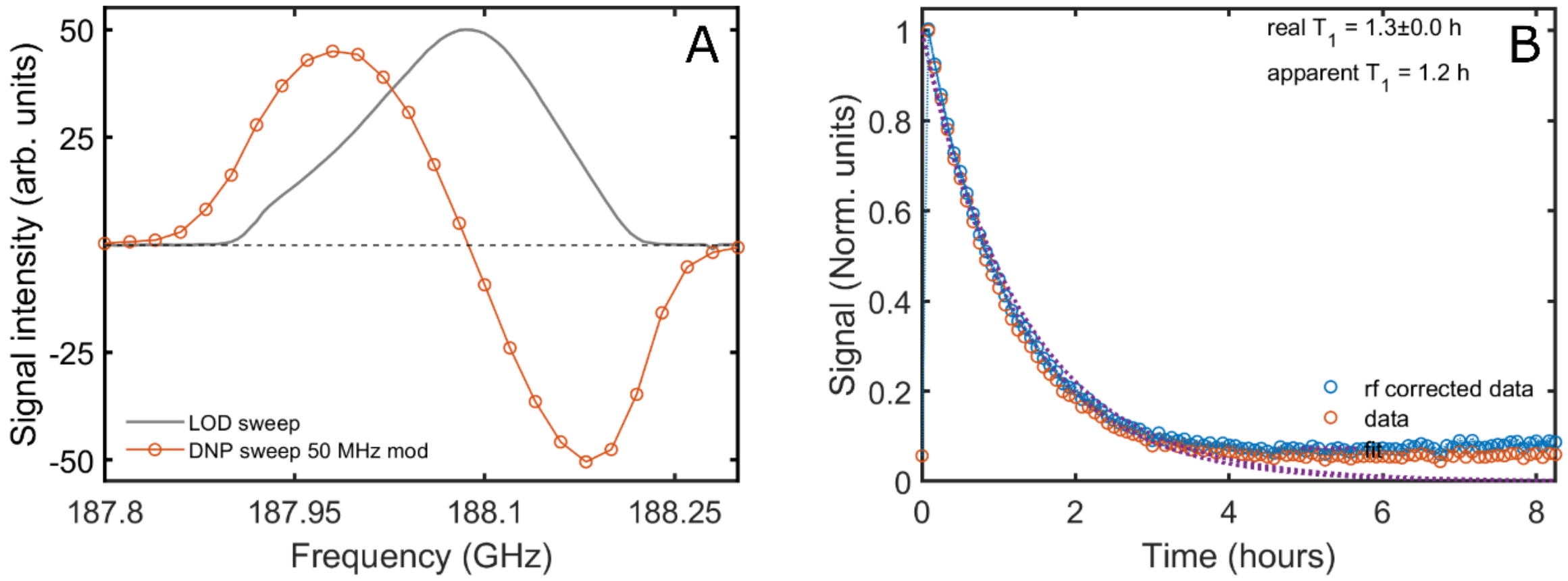

**Figure S12. Note on pyruvic acid:** the DNP performance from the polarizer (**A**) and relaxation from the transportable cryostat,1 T/4.2 K, after thermalization (**B**) are reported for a UV-irradiated sample prepared with 7 M [$d_4$, 1-$^{13}$C]pyruvic acid dissolved in $d_8$-glycerol:$D_2O$.

# References

1. Capozzi, A., Karlsson, M., Petersen, J. R., Lerche, M. H. & Ardenkjaer-Larsen, J. H. Liquid-State $^{13}$C Polarization of 30% through Photoinduced Nonpersistent Radicals. *J. Phys. Chem. C* **122**, 7432–7443 (2018).